\documentclass[12pt,epsf]{article}
\usepackage{graphicx,float}
\usepackage{mathrsfs,array,multirow}
\usepackage{amssymb,amsfonts}
\usepackage{amsmath}
\usepackage{color}
\usepackage{mathrsfs}
\usepackage{amsmath}
\usepackage{amssymb}
\usepackage{amstext}
\usepackage{color}
\newcommand{\be}{\begin{equation}}

\newcommand{\ee}{\end{equation}}

\newcommand{\ba}{\begin{array}}

\newcommand{\ea}{\end{array}}

\begin{document}
\begin{titlepage}

\hfill\parbox{5cm} { }

\vspace{25mm}

\begin{center}
{\Large \bf Conductivities in an anisotropic medium }

\vskip 1. cm
  {Sunly Khimphun$^{a,}$\footnote{e-mail : kpslourk@yahoo.com},
  Bum-Hoon Lee$^{a,b,}$\footnote{e-mail : bhl@sogang.ac.kr} 
  and Chanyong Park$^{b,c,}$\footnote{e-mail : chanyong.park@apctp.org}}

\vskip 0.5cm

{\it $^a\,$Department of Physics, Sogang University, Seoul, Korea 121-742} \\
{\it $^b\,$ Asia Pacific Center for Theoretical Physics, Pohang, 790-784, Korea } \\
{\it $^c\,$ Department of Physics, Postech, Pohang, 790-784, Korea }\\

\end{center}

\thispagestyle{empty}

\vskip2cm


\centerline{\bf ABSTRACT} \vskip 4mm

In order to imitate anisotropic medium of a condensed matter system, we take into account an Einstein-Maxwell-dilaton-axion model as a dual gravity theory where the anisotropy is caused by different momentum relaxations. This gravity model allows an anisotropic charged black hole solution. On this background, we investigate how the linear responses of vector modes like electric, thermoelectric, and thermal conductivities rely on the anisotropy. We find that the electric conductivity in low frequency limit shows a Drude peak and that in the intermediate frequency regime it reveals the power law behavior. Especially, when the anisotropy increases the exponent of the power law becomes smaller. In addition, we find that there exist a critical value for the anisotropy at which the DC conductivity reaches to its maximum value.

\vspace{1cm}

\vspace{2cm}


\end{titlepage}

\renewcommand{\thefootnote}{\arabic{footnote}}
\setcounter{footnote}{0}
\newpage
\tableofcontents

\section{Introduction}

Recently, considerable attention has been paid to the AdS/CFT correspondence (or holography) for understanding strongly interacting systems of nuclear and condensed matter physics. In the strong coupling regime, the perturbation method which is the well-established traditional one does not work, so new physical concepts and/or mathematical techniques are required to figure out strongly interacting systems. In this situation, the AdS/CFT correspondence provides a new way to investigate it nonperturbatively. The AdS/CFT correspondence says that nonperturbative properties of a strongly interacting system can be described by a classical gravity theory defined on an asymptotic AdS geometry \cite{Maldacena:1997re,Witten:1998qj,Witten:1998zw,Gubser:1998bc}, which has already passed many nontrivial tests. In condensed matter physics describing a strongly interacting many-body system, there are many nontrivial important properties like a high $T_c$ superconductivity and scaling behaviors of transport coefficients depending on the phase of matter \cite{Hartnoll:2008kx,Iqbal:2011ae,Hartnoll:2008vx,Huijse:2011ef}. Understanding such properties theoretically, though it is not easy, is one of longstanding problems in physics. After the AdS/CFT correspondence conjecture, there have been numerous attempts to resolve these issues via holographic techniques. In this work, we will study holographically various conductivities of a strongly interacting anisotropic medium and investigate how the anisotropy affects them.

When studying transport coefficients by using the holographic method, it has been well known that, if there exists a translation symmetry, the electric conductivity shows a delta function behavior in the zero frequency limit. This fact means that the DC conductivity is not well defined in a system with a translational symmetry. To resolve this problem, many ideas to break the translational invariance of the dual gravity have been invented. The first is to encode the lattice structure to the AdS geometry which was done by introducing a periodic potential along spatial directions \cite{Horowitz:2012ky,Horowitz:2012gs, Horowitz:2013jaa}. Though this construction can describe the lattice structure of the dual theory, studying its transport coefficients is not easy due to the complexity of the model. Other way to consider the lattice structure is to impose the periodic boundary condition on the chemical potential which is dual to a local gauge field of a dual gravity \cite{Ling:2013nxa, Chesler:2013qla,Donos:2014cya,Donos:2014yya}. One can also take into account a massive gravity theory to break a diffeomorphism invariance \cite{Vegh:2013sk,Blake:2013bqa,Davison:2013jba, Blake:2013owa}. Another way to break the translational symmetry is introducing additional fields depending on spatial coordinates \cite{Koga:2014hwa, Kim:2015wba}. If we introduce scalar fields depending on spatial coordinates linearly, it can mimic the local point of the previous lattice structure. However, since the last method provides a relatively simple calculation it would be a good toy model to understand the transport coefficients of the dual condensed matter system. From now on, we will focus on the last model.

In condensed matter and particle accelerator experiments, anisotropy is one of the important ingredients to understand their physics. There were plenty of works related to the temporal and spatial anisotropies \cite{Koga:2014hwa}-\cite{ Bai:2014poa}. The spatial anisotropy naturally appear by breaking the rotational symmetry of the system. In the last model, it can be easily accomplished by taking different momentum relaxation parameters in $x$- and $y$-directions. On the gravity side, it corresponds to introduce anisotropic axion fields. 
The gravity we will consider has local gauge field, dilaton and axions. In the AdS/CFT contexts, the local bulk gauge field can be identified with the matter with the corresponding global symmetry, while the dilaton field is mapped to the coupling constant of the dual field theory. From these facts, one can expect that the dual field theory of the above gravity describes a medium composed of strongly interacting matter. Adding anisotropic axions to this system, the dual system is modified into an anisotropic medium with different momentum relaxations. In order to clarify properties of the anisotropic medium, we study various conductivities by turning on vector fluctuations on this background geometry. We investigate how the anisotropy affects the transport coefficient like various conductivities.

The rest of the paper is organized as follows: In Sec.\ref{Model}, we construct a dual charged black hole geometry of an anisotropic medium with different momentum relaxations. After turning on vector fluctuations on this background, we investigate the linear responses of the anisotropic medium in Sec. \ref{Conductivity}. We finish this work with some concluding remarks in Sec. \ref{conclusion}.

\section{Einstein-Maxwell-dilaton-axion Model  \label{Model} }
In order to study holographic linear response theory with an anisotropy, let us consider the following action 
\begin{align}\label{Action}
S= \int d^4 x \sqrt{- g}\left(R +\dfrac{6}{L^2}-2 (\nabla\phi)^2-\dfrac{1}{2}e^{ 4\phi}\sum_{i=1}^2 (\nabla \tilde{a}_i)^2 -e^{-2\phi} F^2
\right)\, ,
\end{align}
where $\phi$ and $\tilde{a}_i$ represent a dilaton and two axion fields and $L$ is the AdS radius. 
Following the AdS/CFT correspondence, the profile of the dilaton field can be reinterpreted as a nontrivial running scaling \cite{ Goldstein:2010aw,Kulkarni:2012re,Park:2013ana,Kulkarni:2012in,Park:2013dqa,Park:2014raa}, while the bulk gauge field describes a certain fermionic matter like quark \cite{Lee:2009bya}. This means that the gravity theory we consider mimics a nonconformal medium composed of fermions and gauge bosons \cite{Park:2011zp,Lee:2013oya}. If we introduce the momentum relaxation with an anisotropy by turning on axion fields, the dual field theory can further represent an isotropic medium we are interested in.
In order to describe a momentum relaxation, let us focus on linear axion fields 
\begin{align}
\tilde{a}_1=\alpha_1 x\qquad  \tilde{a}_2=\alpha_2 y  ,
\end{align}
where $\alpha_1$ and $\alpha_2$ are free parameters denoting the momentum relaxation. 
It is well known that, if there is no such axion fields, a DC conductivity generally diverges due to the translational symmetry. On the other hand, a momentum relaxation breaks the translational symmetry and leads to a finite DC 
conductivity. Thus, the existence of axion field plays an important role in studying transport coefficient like a DC conductivity. For $\alpha_1\ne \alpha_2$,
the rotational symmetry is broken, while it is restored only when $\alpha_1=\alpha_2$.  
For the isotropic case, various transport coefficients have been widely investigated (see \cite{Hartnoll:2008kx,Iqbal:2011ae,Hartnoll:2008vx,Herzog:2009xv,Hartnoll:2009sz} and references therein).
However, many samples in condensed matter experiments show an anisotropy. Therefore, it would be interesting to realize such an anisotropy ($\Delta \alpha \equiv  |\alpha_1 - \alpha_2| $) in a holographic model and to investigate its properties.


Equations of motion governing bulk fields are given by 
\begin{align}
& R_{\mu\nu}= -\dfrac{3}{L^2}g_{\mu\nu}+2 \nabla_\mu\phi\nabla_\nu\phi+\dfrac{1}{2}e^{4\phi}\nabla_\mu \tilde{a}\nabla_\nu \tilde{a}+2 e^{-2 \phi}F_{\mu\rho}F_\nu\,^\rho-\dfrac{1}{2}g_{\mu\nu}e^{-2\phi}F^2\,,\label{Eq:Einstein}   \\
&\square \phi -\frac{1}{2} e^{4\phi}\sum_{i=1}^2 (\nabla \tilde{a}_i)^2+\dfrac{1}{2}e^{-2\phi} F^2=0\,, \label{Eq:Dilaton}\\
&\square \tilde{a}_i+4\nabla_{\mu}\phi\nabla^{\mu} \tilde{a}_i=0\,,\label{Eq:Axioni}\\
&\nabla_{\mu}(e^{-2\phi}F^{\mu\nu})=0\,\label{Eq:Maxwell}.
\end{align}
When a rotational symmetry is broken, a general metric ansatz for a black hole has the following form \cite{Iizuka:2012wt,Hartnoll:2009sz}
\begin{align}\label{Metric}
ds^2= \dfrac{L^2}{z^2}\left(-g(z)dt^2 +g(z)^{-1}dz^2+e^{A(z)+B(z)}dx^2+e^{A(z)-B(z)}dy^2\right)\,.
\end{align}
Assuming that the dilaton and the time component of the gauge field are functions of the radial coordinate only
\begin{align}
\phi=\phi(z) , \, {\rm and} \quad   A_\mu dx^\mu= A_t (z) dt,
\end{align}
Eq.(\ref{Eq:Maxwell}) yields
\begin{align}
F_{zt} =A'_{t}= \rho_z L e^{-A+2\phi}\,,
\end{align}
where $\rho_z$ indicates a conserved charge density. Substituting this solution into the other equations, the remaining variables, $A(z)$, $B(z)$ and $\phi (z)$, are governed by
\begin{align}
 2 A''+\left(A'\right)^2+\left(B'\right)^2+4 \left(\phi '\right)^2 &=0\,, \nonumber \\ 
  g z B''+ \left(g \left(z A'-2\right)+z g'\right)B'+\frac{1}{2} z e^{-A-B+4 \phi } \left(\alpha_1 ^2-\alpha_2 ^2 e^{2 B}\right)    &=0\,,    \nonumber              \\
 \left(4 z-2 z^2 A'\right) g'+  \left(-z^2 \left(A'\right)^2+  8 z A'+z^2 \left(B'\right)^2+4 z^2 \left(\phi '\right)^2-12\right)g \nonumber \\
  -\alpha_1 ^2 z^2 e^{-A-B+4 \phi }-\alpha_2 ^2 z^2 e^{-A+B+4 \phi }-4 \rho_z^2 z^4 e^{2 \phi -2 A}+12  &=0\,,    \nonumber  \\
 e^A g \phi ''+ \left(e^A g A'+e^A g'-\frac{2 e^A g}{z}\right)\phi '-\rho_z^2 z^2 e^{2 \phi -A}-\frac{1}{2} \alpha_1 ^2 e^{4 \phi -B}-\frac{1}{2} \alpha_2 ^2 e^{B+4 \phi }&=0\,.\label{EOM:Dilaton}
\end{align}
Note that the equation governing dynamics of $g(z)$ is not independent.

Before solving these equations, it is worth noting that the action we considered is invariant under the following field redefinition 
\begin{align}\label{ScalingDilaton}
\phi\rightarrow\phi-\phi_0\,,\qquad \tilde{a}_1 \rightarrow e^{2\phi_0} \tilde{a}_1\,,\qquad \tilde{a}_2 \rightarrow e^{2\phi_0}\tilde{a}_2 \, , \ {\rm and } \quad   A_t \rightarrow e^{- \phi_0} A_t
\end{align}
where $\phi_0$ implies a constant shift of the dilaton field. Additionally, the metric we have chosen is invariant under the following global scaling 
\begin{align}\label{ScalingAdS}
& A(z) \rightarrow A(z) + A(0) \ , \quad B(z) \rightarrow B(z) + B(0)  \ , \\ \nonumber
& x \rightarrow e^{-(A(0)+B(0))/2}x \, \ {\rm and}  \quad y\rightarrow  e^{-(A(0)-B(0))/2}y ,
\end{align}
where $A(0)$ and $B(0)$ mean the boundary values of $A(z) $ and $B(z)$ respectively.
These three constant shifts imply that one can take arbitrary values of $A$, $B$ and $\phi$ at a given $z$-position, either the horizon or the asymptotic boundary.
In order to see that, let us first introducing a dimensionless 
coordinate scaled by the black hole horizon, $\tilde{z}  = z/z_h$. Then, the black hole horizon appears at $\tilde{z}=1$. 
From now on we always use the $\tilde{z}$ coordinate, so we drop the tilde out for simplicity.
Due to constant shifts of variables, we can set without loss of generality 
\begin{align}\label{HorizonValue1}
A(1)=B(1)=\phi(1)=0\,.
\end{align}
Note that the black hole factor should vanish at the horizon
\begin{align}\label{HorizonValue2}
g(1)=0\,.
\end{align}
Substituting these values into equations of motion,  the first derivatives of variables at the horizon must satisfy the following relations
\begin{align}   \label{res:der}
A'(1)=-\dfrac{12-4 \rho_z^2-\alpha_1^2-\alpha_2^2-4\kappa}{2\kappa}\,,\quad B'(1)=\dfrac{\alpha_1^2-\alpha_2^2}{2\kappa}\,,\quad \phi'(1)=\dfrac{-2 \rho_z^2-\alpha_1^2-\alpha_2^2}{2\kappa}\, ,
\end{align}
where we used $g'(1)=-\kappa$ and $\kappa$ is associated with the Hawking temperature, $\kappa = 4 \pi T$.
Note that since derivatives in (\ref{res:der}) are independent of above constant shifts, 
they still hold even when taking different values from (\ref{HorizonValue1}). This fact plays an crucial role in finding an asymptotic AdS geometry numerically.



\begin{figure}
\vspace{-0.1 cm}
\centering
\includegraphics[width =0.4\textwidth]{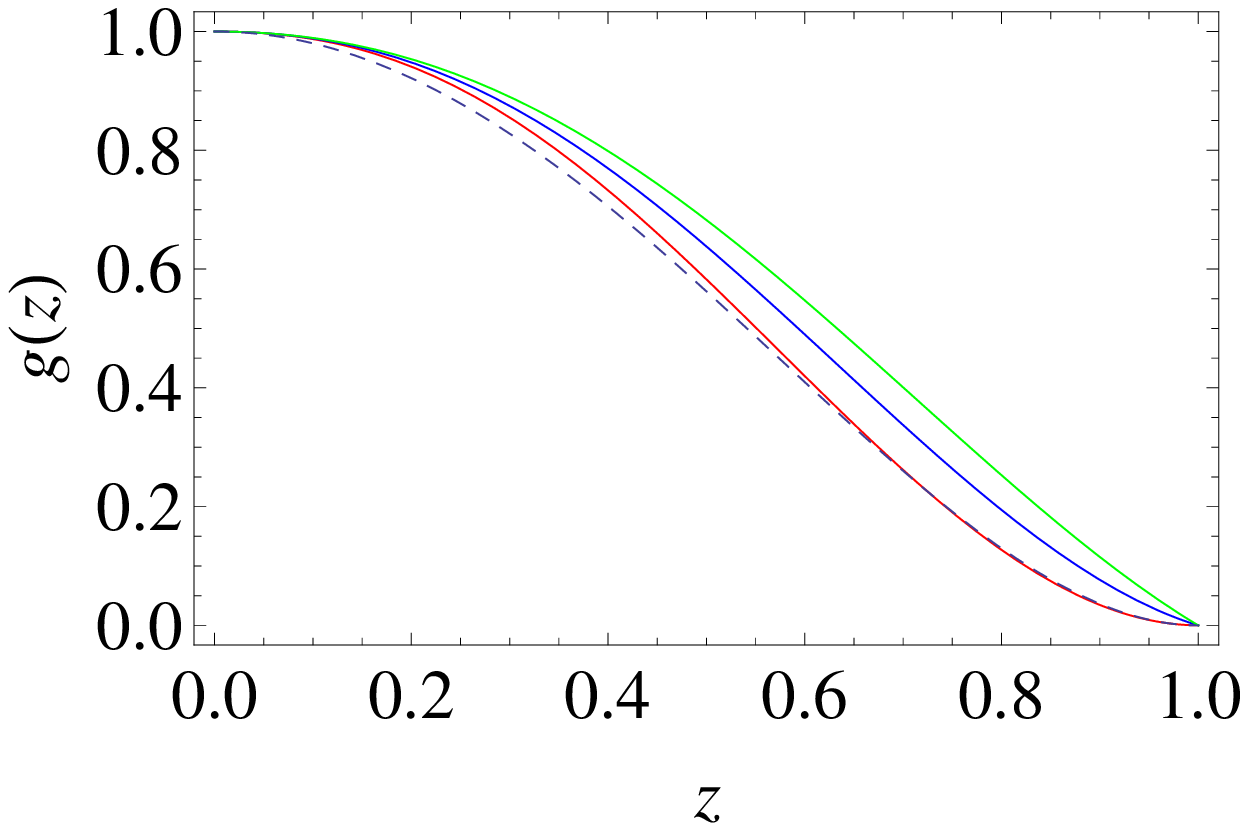}
\hspace{1cm}\includegraphics[width =0.4\textwidth]{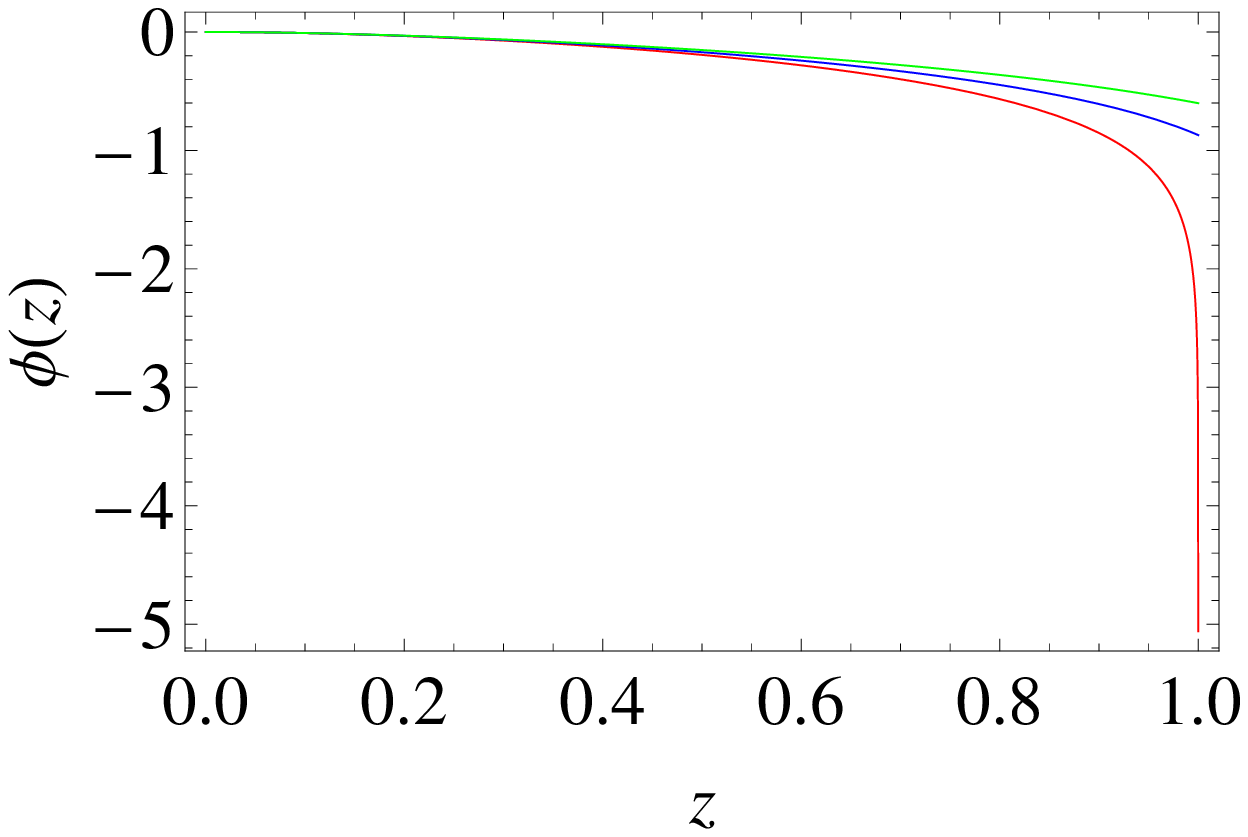}
\includegraphics[width =0.4\textwidth]{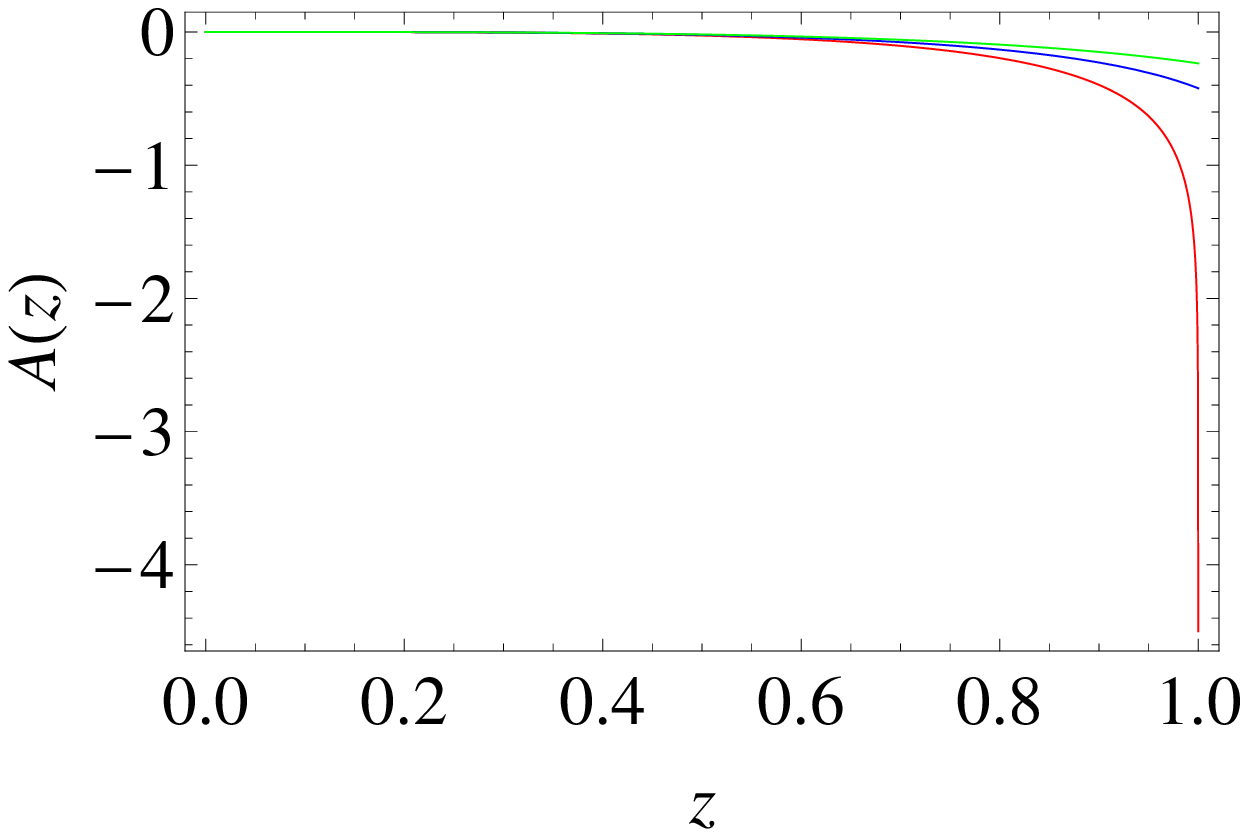}
\hspace{1cm}\includegraphics[width =0.4\textwidth]{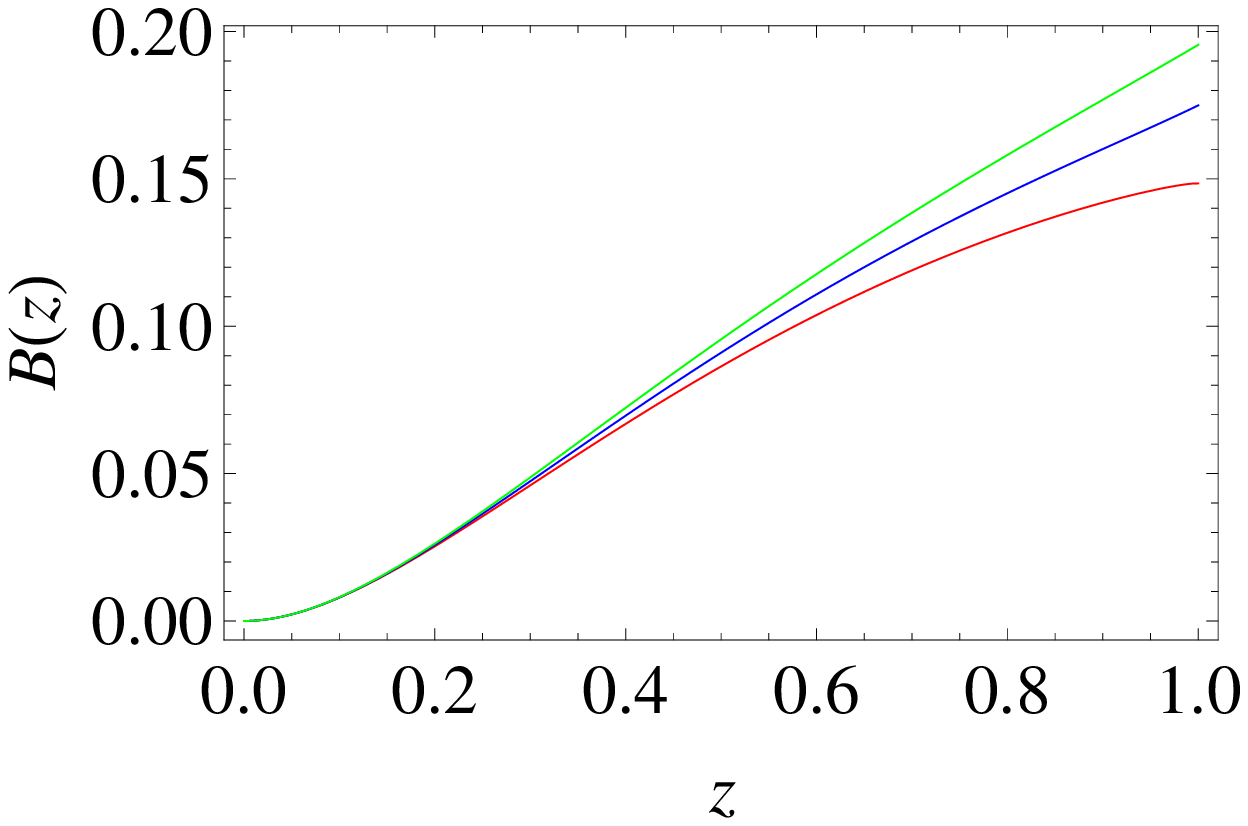}
\vspace{-0.5 cm}
\caption{ Fix $\alpha_1=2$, $\alpha_2=2$, $\kappa=2.4\times 10^{-4}$ (Red), $0.5$ (Blue), and  $1$ (Green). Notice that the red-dashed line is $f(z)$ in (\ref{AdSRNBackSolution}). }\label{F:BackEvolution}
\end{figure}

\begin{figure}
\centering
\includegraphics[width =0.4\textwidth]{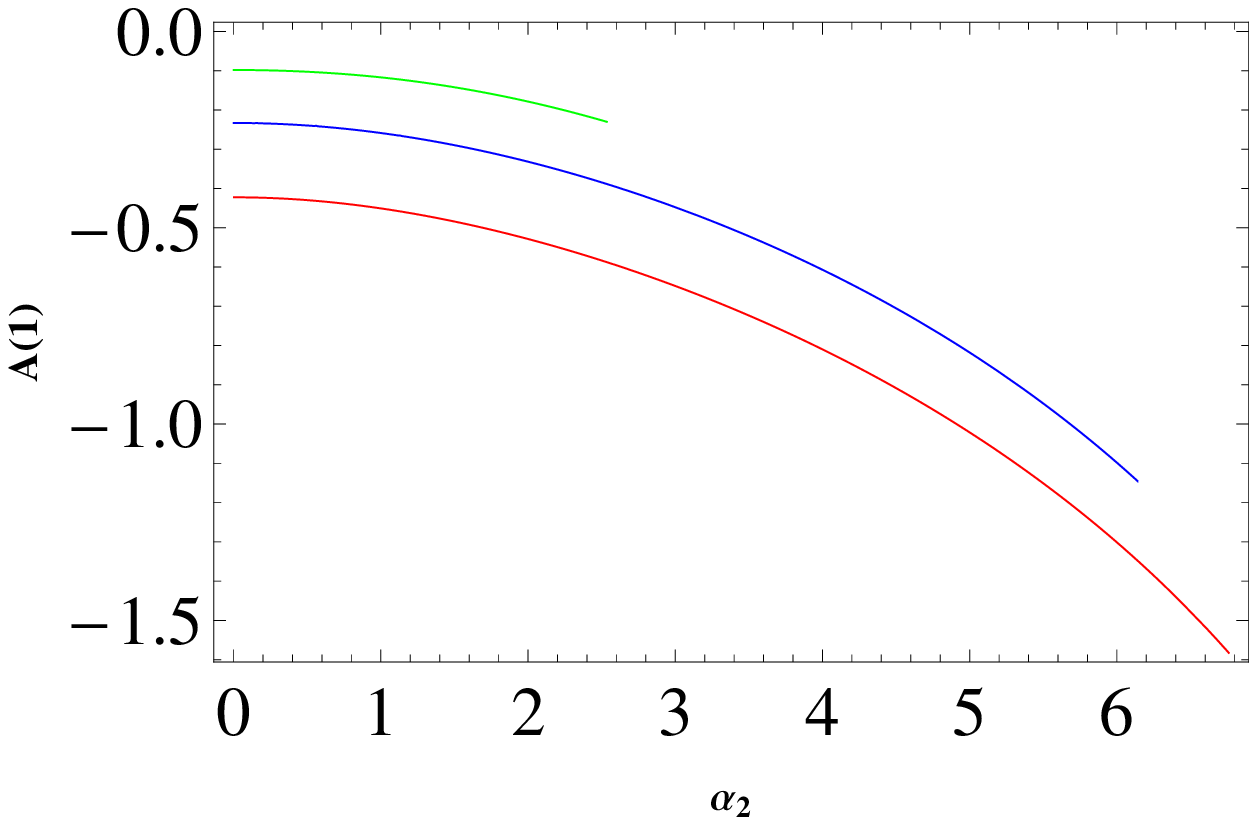}
\hspace{1cm}\includegraphics[width =0.4\textwidth]{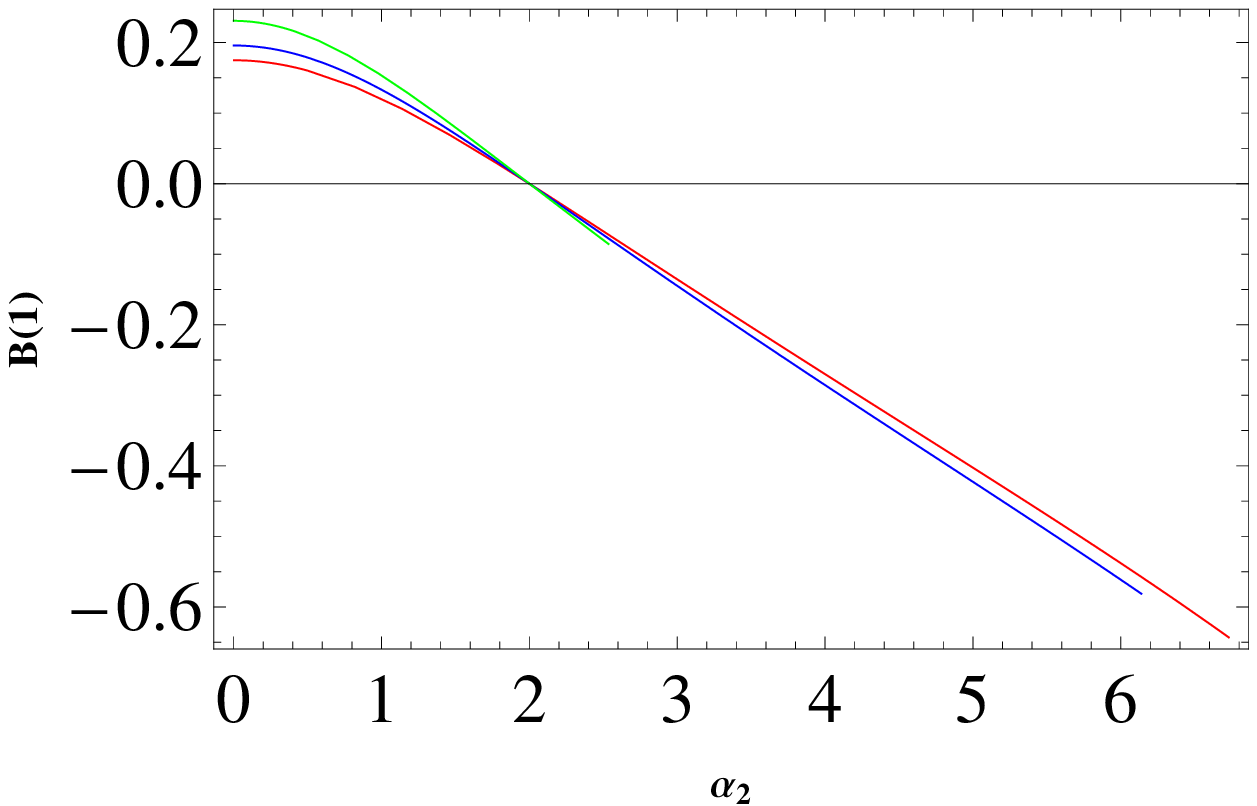}
\includegraphics[width =0.4\textwidth]{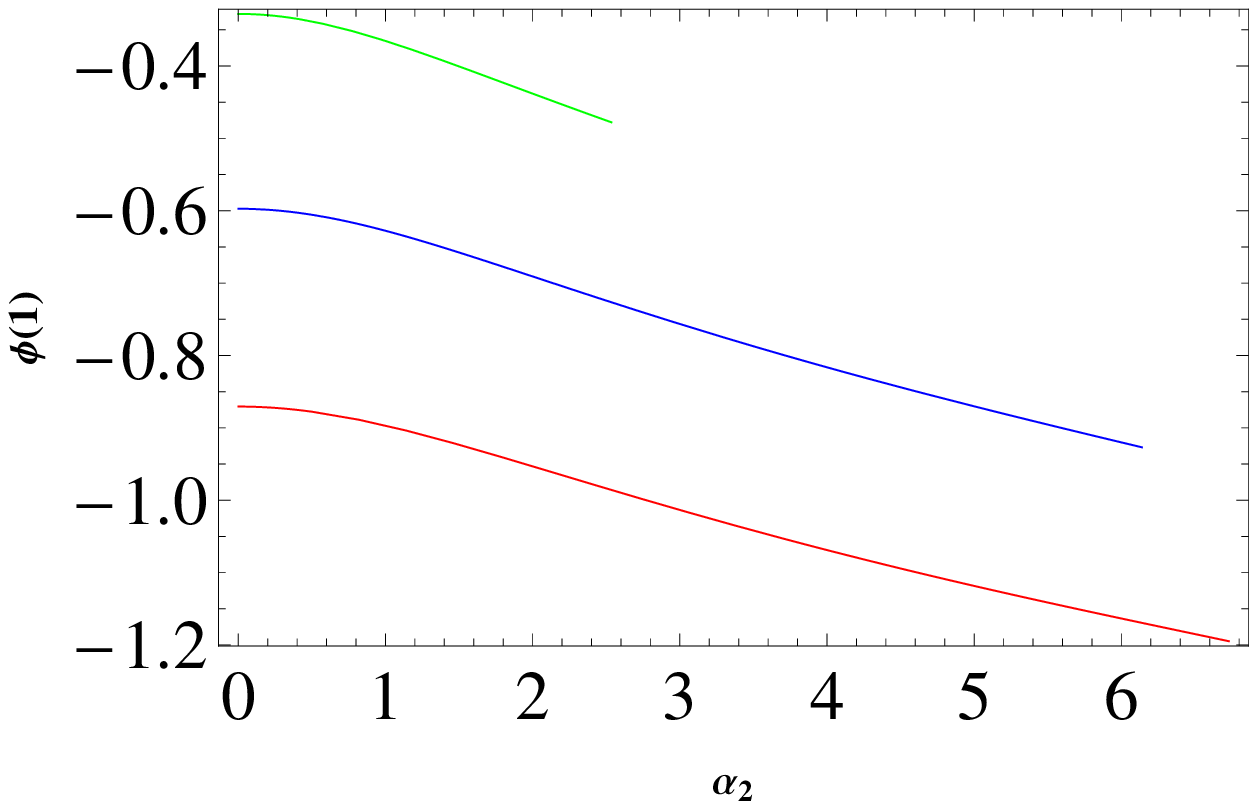}
\hspace{1cm}\includegraphics[width =0.4\textwidth]{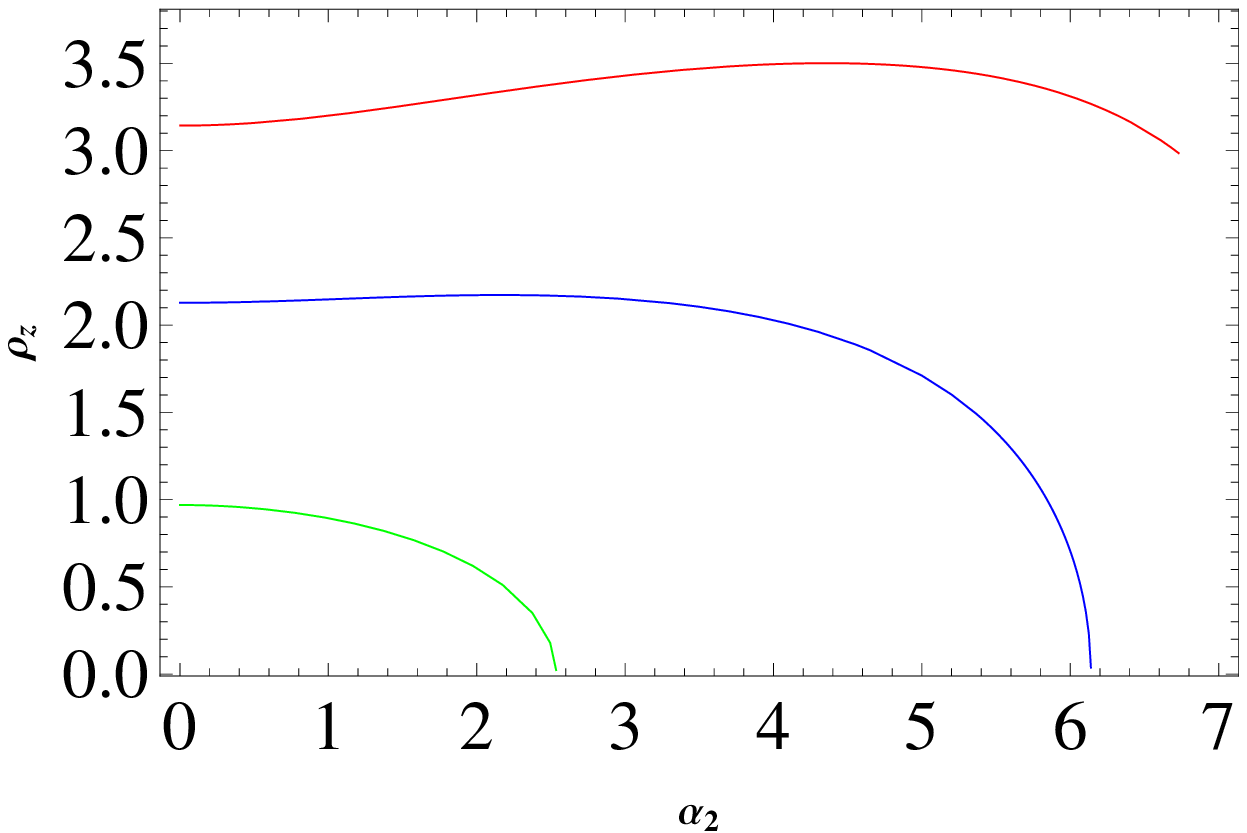}
\vspace{-0.5 cm}
\caption{ Fix $\alpha_1=2$, and $\kappa=0.5$ (Red), $1$ (Blue)  and $2$ (Green)}\label{F:HorizonVSBeta}
\end{figure}

In order to obtain an asymptotic AdS geometry, we should take at the asymptotic boundary
\begin{align}  \label{cond:new}
A(0)=B(0)=0 \   {\rm and} \quad g(0)=1 .
\end{align}
In general, this condition is not consistent with the previous condition in (\ref{HorizonValue1}) defined at the horizon. This is because there is no numerical solution interpolating these two kinds of boundary condition. This fact implies that we have to modify one of boundary conditions, for instance (\ref{HorizonValue1}). Due to the global shift symmetry explained before, it is also allowed to take arbitrary constant values instead of (\ref{HorizonValue1}) which does not give any effect on  (\ref{res:der}). By solving equations of motion in (\ref{EOM:Dilaton}) together with  (\ref{res:der}) and (\ref{cond:new}), it is possible to find consistent values at the horizon which allow continuous interpolation. 
When $\alpha_i$ and $\rho_z$ are given numerical results are depicted in Figure \ref{F:BackEvolution} and \ref{F:HorizonVSBeta}, where we set $L=1$ for convenience.
When parameters are given, we plot profiles of variables in Figure \ref{F:BackEvolution}.  As expected, the boundary values, $A(0) = B(0) = 0$ and $g(0)=1$, indicate that the asymptotic geometry is an AdS space. 

If turning off all scalar fields, the gravity theory we considered reduces to a Reisner-Nordstr\"{o}m AdS black hole which has no scalar hair and anisotropy. If turning off the dilaton field only, this theory for $\alpha_1=\alpha_2=\alpha$ allows the following analytic solution \cite{Kim:2015sma} 
\begin{align}\label{AdSRNBackSolution}
ds^2= \dfrac{L^2}{z^2}\left(-f(z)dt^2 +f(z)^{-1}dz^2+dx^2+dy^2\right)\,,
\end{align}  
with
\begin{align}
f(z) &= 1-\frac{\alpha ^2}{2}  z^2 - 
\left(1 - \frac{\alpha ^2}{2}+\frac{\mu ^2}{4   L^2}\right) z^3 + \frac{\mu ^2 }{4 L^2}  z^4 \ .
\end{align}

In Figure \ref{F:HorizonVSBeta},  we depict the horizon values relying on parameters. The magnitudes of $A(1)$ and $\phi(1)$ decrease as $\alpha_2$ increases, while they increases as the temperature increases. Especially for $\alpha_1 = \alpha_2 = 2$, $B(1)$ vanishes because the isotropy is restored at this point. Another interesting point we should notice is that when the temperature increases $B(1)$ increases for $\alpha_2 < \alpha_1$, whereas decreases for $\alpha_2>\alpha_1$. In our work $\alpha_i$ and $\kappa$ are chosen as independent parameters, so $\rho_z$ depends on these parameters. Following the AdS/CFT correspondence, the geometry we found numerically can be reinterpreted as an anisotropic medium on the dual field theory side. In this case, the anisotropy is caused by the different momentum relaxation in $x$- and $y$-directions.

Before closing this section, let us discuss on the perturbative solution near the boundary. Denoting bulk fields as $\Pi$ collectively, they can be expanded into the following form near the boundary
\begin{align}\label{BoundaryField}
\Pi= \lim_{z\rightarrow0}\sum_{n=0}^\infty \Pi^{(n)}z^{n}\,.
\end{align}
the perturbative solution satisfying equations of motion in (\ref{EOM:Dilaton}) are given by
\begin{align}\label{dependentCoeffientBack}
 A^{(0)} &=B^{(0)}=\phi^{(0)}=0,\qquad  g^{(0)} =1,\qquad  B^{(1)} =\phi^{(1)}=A^{(1)}=g^{(1)}=0,\nonumber\\
 A^{(2)} &=A^{(3)}=0,\qquad  \phi^{(2)}  =g^{(2)}=-\frac{1}{4}(\alpha_1^2+\alpha_2^2),\qquad  B^{(2)} =\frac{1}{4}(\alpha_1^2-\alpha_2^2),\nonumber\\
 A^{(4)}  &= -\frac{1}{96}(5 \alpha_1^4+6 \alpha_1^2\alpha_2^2+5 \alpha_2^4),\qquad \phi^{(4)} = -\frac{1}{16}(3 \alpha_1^4+4 \alpha_1^2\alpha_2^2+3 \alpha_2^4-4 \rho_z^2), \nonumber\\
g^{(4)}&= \frac{1}{24}(24\rho_z^2-5\alpha_1^4-6\alpha_1^2\alpha_2^2-5\alpha_2^4) ,
\nonumber\\
 B^{(4)} &= \frac{3}{16}\left( \alpha_1^2 -\alpha_2^2 \right) \, , 
 \nonumber \\
& \quad \cdots .
\end{align}
Above $g^{(3)}$, $B^{(3)}$, $\phi^{(3)}$ are undetermined integral constants which can be fixed by imposing additional boundary conditions at the horizon. This perturbative solution plays an important role in calculating the on-shell gravity action and in evaluating conductivities of the dual anisotropic medium.

\section{Conductivities in an anisotropic medium}\label{Conductivity}

In order to investigate various conductivities of the anisotropic medium considered in the previous section, we perturb the gauge field 
\begin{align*}
A_\mu dx^\mu \rightarrow A_t(z) dt +\left[\tilde{A}_x(t,z) dx +\tilde{A}_y(t,z) dy \right]\,,
\end{align*}
with the metric fluctuation $\tilde{g}_{ti}$ $(i,j=x,y)$
\begin{align*}
g_{\mu \nu} dx^{\mu} dx^\nu \rightarrow  \overline{g}_{\mu \nu} dx^\mu dx^\nu+
\dfrac{2 L^2}{z^2}
\left[\tilde{g}_{tx}(t,z) dtdx  +\tilde{g}_{ty}(t,z) dtdy \right]\,,
\end{align*}
where $\overline{g}_{\mu \nu}$ is the background metric we found in the previous section. 
The other metric fluctuations like
$g_{ij}$ are not considered here because they are usually decoupled in the linear response theory \cite{Policastro:2002se,Kovtun:2012rj,Kim:2015sma}. Note that, since fluctuations of the axion field can be coupled to above gauge field and metric fluctuations, one should also take into account axion's fluctuation. 

Now, let us consider the following Fourier mode decompositions
\begin{align*}
\tilde{A}_i(t,z)= \int_{-\infty}^{\infty} \frac{d\Omega}{2\pi}e^{-i \Omega t}A_i(z),\qquad
\tilde{g}_{ti}(t,z)= \int_{-\infty}^{\infty} \frac{d\Omega}{2\pi} e^{-i \Omega t}g_{ti}(z),
\end{align*}
and
\begin{align}
\tilde{a}_i \rightarrow \alpha_i x^i +i \int_{-\infty}^{\infty} \frac{d\Omega}{2\pi} \Omega e^{-i \Omega t} \chi_i (z)\, ,
\end{align}
where $\Omega$ is a dimensionless frequency defined as $\Omega =\dfrac{L^2}{r_{h}} \omega$ \cite{Koga:2014hwa}. 
Then, fluctuations are governed by the following equations of motion
\begin{align}
A_i''+ \left(\bar{B_i}'+\frac{g'}{g}-2 \phi '\right)A_i'+\left(\frac{\Omega ^2}{g^2}-
\frac{4 z^2 e^{2 \phi -2 A} \rho _z^2}{g}\right)A_i -\alpha_i  L e^{6 \phi -A}  
\rho _z \chi_i' &=0\,,\label{Peq:Ai}\\
\chi _i''+ \left(A'+\frac{g'}{g}-\frac{2}{z}+4 \phi '\right)\chi _i'+\frac{ \Omega ^2}{g^2}\chi _i-\frac{\alpha_i  e^{\bar{B_i}-A}}{g^2} g_{\text{ti}}&=0\,,\label{Peq:chii}\\
g_{ti}'+ \left(\bar{B_i}'-A'\right)g_{ti}-\frac{4 e^{-A} z^2  \rho _z}{L}A_i-\alpha_i  g e^{4 \phi } \chi _i'&=0\,.\label{Peq:gti}
\end{align}
where $\bar{B_i}=\left\{ -B,B \right\}$ for $i=\left\{ x,y \right\}$ respectively. In the above equation, the fluctuations in 
$x$- and $y$-directions seem to be decoupled. If it is true, we can consider only the fluctuation in $x$-direction because 
the fluctuation in $y$-direction follows the same equation of motion just with a different momentum relaxation parameter.
However, this is not true in the present work. Since the background fields, $A$, $B$, $g$ and $\phi$, are functions of both relaxation parameters, $\alpha_x$ and $\alpha_y$,  the fluctuations in $x$- and $y$-directions are implicitly coupled. Therefore, the momentum relaxation parameter in $x$-direction can affect conductivities in $y$-direction as well as $x$-direction.
This implies that the anisotropic medium we considered depends nontrivially on the momentum relaxations.

In order to solve above equations, let us first focus on the near horizon behavior of solutions.
Above equations (\ref{Peq:Ai})-(\ref{Peq:gti}) has a singularity at the horizon due to $g(1)=0$. For a well-defined solution at the horizon, fluctuations must show appropriate singular behaviors at the horizon. 
Introducing new variables,  
\begin{align}
\hat{A}_i(z)\equiv g(z) A_i'(z)\,, \quad \hat{\chi_i}(z)\equiv g(z)\chi'_i(z)\,,
\end{align}
and 
\begin{align*}
A_i(z)&=(1-z)^\lambda a_i(z)\,, &  \hat{A}_i(z) &=(1-z)^\lambda\hat{a}_i(z)\,,& g_{ti}(z)& =(1-z)^\lambda\zeta_{ti}(z)\\
\chi_i(z)&=(1-z)^\lambda\eta_i(z)\,,   &  \hat{\chi_i}(z)&=(1-z)^\lambda \hat{\eta_i}(z)\,,
\end{align*}
with an appropriate exponent $\lambda$, the above equations in terms of new variables can be rewritten as five first-order linear differential equations in each direction
\begin{align}
\hat{a}_i'+ \left(\bar{B_i}'-\frac{\lambda }{1-z}-2 \phi '\right)\hat{a}_i+ \left(\frac{\Omega ^2}{g}-4 z^2 e^{2 \phi -2 A} \rho _z^2\right)a_i-\theta_i \rho _z  L e^{6 \phi -A} \hat{\eta }_i &=0\,,\label{PHeq:aih}\\
a_i'-\frac{\lambda}{1-z}a_i-\frac{\hat{a}_i}{g} &=0\,,\label{PHeq:ai}\\
\hat{\eta }_i'+ \left(A'-\frac{\lambda }{1-z}-\frac{2}{z}+4 \phi '\right)\hat{\eta }_i+\frac{ \Omega ^2}{g}\eta _i-\frac{\theta_i  e^{-A+\bar{B_i}} \zeta_{ti}}{g}  &=0\,,\label{PHeq:Etaih}  \\
\eta _i'-\frac{ \lambda }{1-z}\eta _i-\frac{\hat{\eta }_i}{g}&=0\,,\label{PHeq:Etai}\\
\zeta_{ti}'+ \left(-A'+\bar{B_i}'-\frac{\lambda }{1-z}\right)\zeta_{ti}-\theta_i  e^{4 \phi }\hat{\eta }_i -\frac{4 \rho _z e^{-A} z^2 a_i }{L} &=0\,.\label{PHeq:Zetati}
\end{align}
At the horizon, these equations lead to the following eigenvalue equation
\begin{align}
 \left(
\begin{array}{ccccc}
 0  & -\frac{\Omega ^2}{g'(1)} & 0 & 0 & 0 \\
 \frac{1}{g'(1)} & 0  & 0 & 0 & 0 \\
 0 & 0 & 0  & -\frac{\Omega ^2}{g'(1)} & \frac{ \alpha_i e^{-A(1)-\bar{B_i}(1)} }{g'(1)} \\
 0 & 0 & \frac{1}{g'(1)} & 0  & 0 \\
 0 & 0 & 0 & 0 & 0  \\
\end{array}
\right) \left(
\begin{array}{c}
 \hat{a}_i \\
 a_i \\
 \hat{\eta }_i \\
 \eta _i \\
 \zeta _{\text{ti}} \\
\end{array}
\right)&=\lambda  \left(
\begin{array}{c}
 \hat{a}_i \\
 a_i \\
 \hat{\eta }_i \\
 \eta _i \\
 \zeta _{\text{ti}} \\
\end{array}
\right)\,,\label{EigenEquationX}
\end{align}
In each direction, this eigenvalue equation allows three degenerate eigenvalues, $\lambda=0$ and $\lambda=\pm \frac{i \Omega}{g'(1)}$, whose eigenvectors are given by 

\begin{align}
\psi_0 = \left(
\begin{array}{c}
 0 \\
 0 \\
 0 \\
 \frac{\theta_i  e^{-A(1)+\bar{B}_i(1)}}{\Omega ^2} \\
 1 \\
\end{array}
\right)\,, \quad 
\psi_{1\pm} =\left(
\begin{array}{c}
 0 \\
 0 \\
\pm i \Omega  \\
 1 \\
 0 \\
\end{array}
\right)\,,\quad
\psi_{2\pm} =\left(
\begin{array}{c}
\pm  i \Omega  \\
 1 \\
 0 \\
 0 \\
 0 \\
\end{array}
\right)\, . \label{EigenVectorH}
\end{align}
Here the case with $i \lambda >0$ satisfies an incoming boundary condition, while the solution satisfying the outgoing boundary condition appears for $i \lambda < 0$. Note that $g(z)$ in the outside of the horizon must be positive and becomes zero at the horizon. This fact implies that $g'(1)$ is always negative as shown in Figure \ref{F:BackEvolution}.  
Assuming $\Omega >0$, $\psi_{1+}$ and $\psi_{2+}$ correspond to eigenvectors satisfying the incoming boundary condition whose eigenvalue is given by $\lambda = \frac{i \Omega}{g'(1)}$. Removing outgoing modes at the horizon, $\chi_{i}$, $g_{ti}$ and $A_i$ are determined as the linear combination of incoming modes and zero mode with three coefficients, $ c_0 \psi_0 + c_1 \psi_{1+}+ c_2 \psi_{2+}$. As a consequence, solutions can be uniquely determined by fixing these three coefficients. In order to fix them, we have to impose additional boundary conditions. Imposing Dirichlet boundary conditions on $\chi_{i}$, $g_{ti}$ and $A_i$ at the asymptotic boundary  clarifies their boundary values and at the same time fix above coefficients.

\subsection{Conductivities of an anisotropic medium}

In the previous section, we discussed how we can obtain a numerical fluctuation solution satisfying equations of motion and consistent boundary conditions. In order to extract some physical information following the AdS/CFT correspondence, we need to understand further its structure near the boundary. Since $z \ll 1$ near the boundary, the numerical solution obtained in the previous section allows the following perturbative expansion 
\begin{align}   \label{exp:fluc}
\Phi = \lim_{z\rightarrow0}\sum_{n=0}^\infty \Phi^{(n)}z^{n}\,.
\end{align}
where $\Phi$ indicates all fluctuations like $\chi_i$, $g_{ti}$ and $A_i$ collectively. Since three fluctuations we considered satisfy the second order differential equations, they usually have six integral constants, $\chi_i^{(0)}$, $\chi_i^{(3)}$, $g_{ti}^{(0)}$, $g_{ti}^{(3)}$, $A_i^{(0)}$ and $A_i^{(1)}$. Following the holographic prescription $\chi_i^{(0)}$, $g_{ti}^{(0)}$ and $A_i^{(0)}$ are mapped to sources, while $\chi_i^{(3)}$, $g_{ti}^{(3)}$ and $A_i^{(1)}$ are interpreted as vev of dual operators on the dual field theory side. Especially, $A_i^{(1)}$ and $g_{ti}^{(3)}$ are dual of the electric current, $J^i$, and momentum operator, $T^{ti}$, respectively. Note that because of one constraint equation in the Einstein equation, one of them can be rewritten in terms of the others. As a consequence, only five coefficients are independent and the remaining are usually determined by these five integral constants. More precisely, we can fix $g_{ti}^{(3)}$ in terms of the other five coefficients by solving constraint equation 
\begin{align}\label{DependentCoeffient}
g_{ti}^{(3)}&= \frac{4 A_i^{(0)} \rho_z+3 \alpha_i  L \chi_i^{(3)}}{3 L}\, .
\end{align}
The other coefficients in (\ref{exp:fluc}) can be also fixed by five independent integral constants. Here we present several lower order coefficients 
\begin{align}\label{DependentCoeffient}
\chi_i^{(1)}&=g_{ti}^{(1)}=0 \,, \nonumber\\
\quad \chi_i^{(2)} &=\frac{1}{2} \left(\chi_i^{(0)} \Omega ^2-\alpha_i g_{ti}^{(0)}\right)\,, \qquad g_{ti}^{(2)} = \frac{1}{4} \left(2 \alpha_i  \chi_i^{(0)} \Omega ^2-\alpha_1 ^2  g_{ti}^{(0)}-\alpha_2 ^2 g_{ti}^{(0)}\right)\,, \nonumber\\
 A_i^{(2)} &= \dfrac{-\Omega }{2} A_i^{(0)} \,,  \nonumber\\
A_{i}^{(3)}&= \frac{1}{6} \left(- \alpha_j ^2 A_i^{(1)}- \Omega ^2 A_i^{(1)}- \alpha_i ^2 L \rho _z g_{ti}^{(0)}+ \alpha_i  L \Omega ^2 \rho _z \chi_{i}^{(0)}\right)\,,\nonumber\\
A_{i}^{(4)}&= \frac{1}{24} \left((-1)^{i} (\alpha_1 ^2- \alpha_2 ^2 ) A_i^{(0)} \Omega ^2+8 A_i^{(0)} \rho_z^2-6 A_i^{(1)} g^{(3)}+ A_i^{(0)} \Omega ^4+6 \alpha_i  \rho_z L \chi_i^{(3)}\right)\, ,
\end{align}
where $i\neq j$ and $g^{(3)}$ corresponds to the third order term of the background black hole metric factor, $g(z)$, in (\ref{Metric}). Comparing these perturbative solutions with the previous numerical solution, it is possible to know the exact numerical values of all coefficients. 

Knowing these coefficients exactly is important to understand physical properties of the dual field theory because the on-shell gravity action determined by them plays a role of a generating functional of the dual theory. Since the variation of a gravity action is not well defined, we need to add an additional boundary term called the Gibbons-Hawking term for the well-defined variation. The on-shell gravity action and the Gibbons-Hawking term are given by
\begin{align}
S_{on} &= \int \frac{ d^3 x}{2 \pi}  \left[g_{ti} \left(-4 \rho_z L e^{-A+\bar{B_i}} A_i+\frac{2 e^{\bar{B_i}} L^2 g_{\text{ti}}'}{z^2}-\frac{\alpha_i  g L^2 e^{4 \phi+\bar{B_i}}\chi_{i}'}{2 z^2}\right)\nonumber\right.\\
& \qquad \qquad \left.\quad  +g_{\text{ti}}^2 \left(-\frac{e^{\bar{B_i}} L^2 A'}{z^2}+\frac{e^{\bar{B_i}} L^2 \bar{B'_i}}{z^2}-\frac{e^{\bar{B_i'}} L^2 g'}{2 g z^2}-\frac{e^{\bar{B_i}} L^2}{z^3}\right) \ \right]  ,\\
S_{GH} &=  -2 \int \frac{ d^3 x}{2 \pi} \sqrt{-\gamma} \  K ,
\end{align}
where $\gamma$ denotes an induced metric at the boundary ($z \to 0$). In general, these actions have divergent terms corresponding to UV divergences of the dual field theory. Similar to a usual quantum field theory, it should be renormalized by adding appropriate counter terms which is called a holographic renormalization \cite{de Boer:1999xf,Balasubramanian:1999re,deHaro:2000vlm,Park:2013ana,Park:2013dqa,Park:2014gja}. Proper counter terms removing UV divergences are given by
\begin{align}\label{CounterTerm}
S_{ct}  = \int \frac{ d^3 x}{2 \pi} & \sqrt{-\gamma}\left[-\frac{4}{L}+\frac{L}{2}
\sum_{i=1}^{2}\gamma^{m n}\partial_{m} \tilde{a}_i\partial_{n} \tilde{a}_{i} \right]  .
\end{align}
Substituting perturbative solutions to the fluctuation's action yields the renormalized on-shell gravity action corresponding to a generating functional of the dual field theory  
\begin{align}\label{RenormalizeAction}
S_{re}^{(2)} &=  S_{on} +S_{GH} +S_{ct}  \nonumber\\
& =  2\int d^2 x\int \frac{d \Omega}{2\pi} \left[ A_i^{(0)}A_i^{(1)}+\frac{1}{2}L^2 \left(  p_i B^{(3)}+g^{(3)} \right) g_{ti}^{(0)}g_{ti}^{(0)}-2 L\rho_z  A_{i}^{(0)}g_{ti}^{(0)}  \nonumber\right.\\
&\left.\qquad\qquad\qquad\qquad +\frac{3}{4}L^2\Omega^2\chi_i^{(0)}\chi_i^{(3)} -\frac{3}{4}L^2\alpha_i g_{ti}^{(0)}\chi_{i}^{(3)} \right] \,,
\end{align}
where $p_i$ is either $1$ for $i=x$ or $-1$ for $i=y$. From this finite renormalized action, one can easily extract a retarded Green function following the holographic prescription \cite{Policastro:2001yc,Policastro:2002tn,Policastro:2002tn}.

Near the boundary, fluctuations usually allow two independent solutions 
\begin{align}
\Phi_{i}^{\mathfrak{a}} = {\mathbb{S}_i^\mathfrak{a}} z^{3-\Delta_\mathfrak{a}} +\cdots +\mathbb{O}_i^\mathfrak{a} z^{\Delta_\mathfrak{a}}+\cdots\,,
\end{align}
where different fields, $A_i, g_{ti}$ and $\chi_i$, are distinguished by an index $\mathfrak{a}$. Here 
$\Delta_\mathfrak{a}$ is a positive value and corresponds to the conformal dimension of the dual operator. Then,
the on-shell gravity action can be written as the following form
\begin{align}\label{OnShellAction2}
S_{re} =2V\int\frac{d\Omega}{2\pi}\left[ \bar{\mathbb{S}}_i^\mathfrak{a} \mathbb{A}^{ij}_{\mathfrak{a} \mathfrak{b}}(\Omega) \mathbb{S}_j^\mathfrak{b}
                 +\bar{\mathbb{S}}_i^\mathfrak{a} \mathbb{B}^{ij}_{\mathfrak{a} \mathfrak{b}} (\Omega) \mathbb{O}_j^\mathfrak{b} \right]\,,
\end{align}
where 
\begin{align}
\mathbb{S}_i &\equiv 
\left(\begin{array}{c}
          \mathbb{S}_i^1 \\   \mathbb{S}_i^2 \\    \mathbb{S}_i^3 \\ 
\end{array} \right)  =
\left(\begin{array}{c}
           A_{i}^{(0)} \\ g_{ti}^{(0)}\\  \chi_i^{(0)} \\ 
\end{array} \right) \,, \quad 
\mathbb{O}_i\equiv 
\left(\begin{array}{c}
          \mathbb{O}_i^1 \\   \mathbb{O}_i^2 \\    \mathbb{O}_i^3 \\ 
\end{array} \right)  =
\left(\begin{array}{c}
 A_{i}^{(1)} \\ g_{ti}^{(3)} \\ \chi_i^{(3)}\\ 
\end{array} \right)\,  ,
\end{align}
and $V$ is the regularized spatial volume. Comparing it with the previous renormalized on-shell gravity action, we obtain
\begin{align}\label{ABMatrice}
\mathbb{A}^{ij} = \left(\begin{array}{ccc}
0 & - L \rho_z  & 0\\
- L \rho_z & \frac{1}{4} L^2(g^{(3)}+ p_i B^{(3)})   & 0\\
0 &  0  & 0\\
\end{array} \right) \delta^{ij} \,,\qquad 
\mathbb{B}^{ij} = \left(\begin{array}{ccc}
1   & 0       & 0\\
0   & 0  & -\frac{3 L^2 \alpha_i}{4} \\
0   &  0      & \frac{3 L^2 \Omega^2}{4} \\
\end{array} \right) \delta^{ij} \,,
\end{align}
and the retarded Green function is given by
\begin{align}
G^{ij}_{\mathfrak{a}\mathfrak{b}} \equiv \mathbb{A}^{ij}_{\mathfrak{a}\mathfrak{b}} +\mathbb{B}^{ik}_{\mathfrak{a}\mathfrak{c}} \mathbb{O}_k^\mathfrak{c}{(\mathbb{S}^{-1})}_{\mathfrak{b}}^j   ,
\end{align}
where ${(\mathbb{S}^{-1})}_{\mathfrak{b}}^j$ means $1/{(\mathbb{S})}_{\mathfrak{b}}^j$. If we focus on the gauge and metric fluctuations, the linear response to the variation of the source is given by
\begin{align}
\left(
\begin{array}{c}
 J^{j} \\
T^{tj} \\
\end{array}
\right) = \left(
\begin{array}{ccccc}
G^{ij}_{11} & G^{ij}_{12}\\
G^{ij}_{21} &G^{ij}_{22}\\
\end{array}
\right) \left(
\begin{array}{c}\label{TransportCoef}
A_i^{(0)}\\
g_{ti}^{(0)}\\
\end{array}
\right)\, ,
\end{align}
where we used $ J^{j}=A^{j(1)}$ and $T^{tj}=g^{tj(3)}$.

Now, let us compare this result with a known form of the linear response theory
\cite{Hartnoll:2008kx,Iqbal:2011ae,Hartnoll:2008vx,Herzog:2009xv,Hartnoll:2009sz}
\begin{align}
\left(
\begin{array}{c}
 J^{i} \\
 Q^{i} \\
\end{array}
\right) = \left(
\begin{array}{ccccc}
\sigma & \tilde{\alpha} T\\
\bar{\alpha} T & \bar{\kappa} T\\
\end{array}
\right) \left(
\begin{array}{c}\label{TransportCoef}
E_i\\
-(\nabla_i T)/T\\
\end{array}
\right)\, ,
\end{align}
where $Q^i$ indicates a heat current, $Q^i =T^{ti} - \mu J^i$ with a chemical potential $\mu$ defined by the boundary value of the background gauge field $A_t$.
Note that the source terms of fluctuations can be identified with the electric field and thermal gradient due to the diffeomorphism invariance \cite{Kim:2015wba,Hartnoll:2009sz,Hartnoll:2008hs}
\begin{align}
E_i=i \Omega(A_{i}^{(0)}+\mu g_{ti}^{(0)}) \quad {\rm and} \qquad g_{ti}^{(0)}=-\frac{\nabla_i T}{i \Omega T}\, .
\end{align}
As a consequence, the transport coefficients can be rewritten in terms of the retarded Green functions \cite{Kim:2015wba} 
\begin{align}
\left(\begin{array}{ccc}
\sigma_{ii} & \tilde{\alpha}_{ii} T \\
\bar{\alpha}_{ii} T & \bar{\kappa}_{ii}T  \\
\end{array} \right)= \left(\begin{array}{ccc}
-\frac{i G_{11}^{ii}}{\Omega} & \frac{i(\mu G_{11}^{ii}-G_{12}^{ii})}{\Omega}\\
\frac{i(\mu G_{11}^{ii} - G_{21}^{ii})}{\Omega} & -\frac{i \left[ G_{22}^{ii}- \tilde{G}_{22}^{ii}-\mu(G_{12}^{ii}+G_{21}^{ii}-\mu G_{11}^{ii}) \right]}{\Omega}
\end{array} \right) .
\end{align}
Usually, $G_{22}^{ii}$ does not vanish when $\Omega \to 0$,  If we denote this non-zero value as $\tilde{G}_{22}^{ii}$, it yields a divergence corresponding to a contact term. Above we subtract such a divergence for a well-defined thermal conductivity \cite{Kovtun:2012rj, Kim:2014bza}.

\subsection{Numerical Result}

\begin{figure}
\centering
\includegraphics[width =0.4\textwidth]{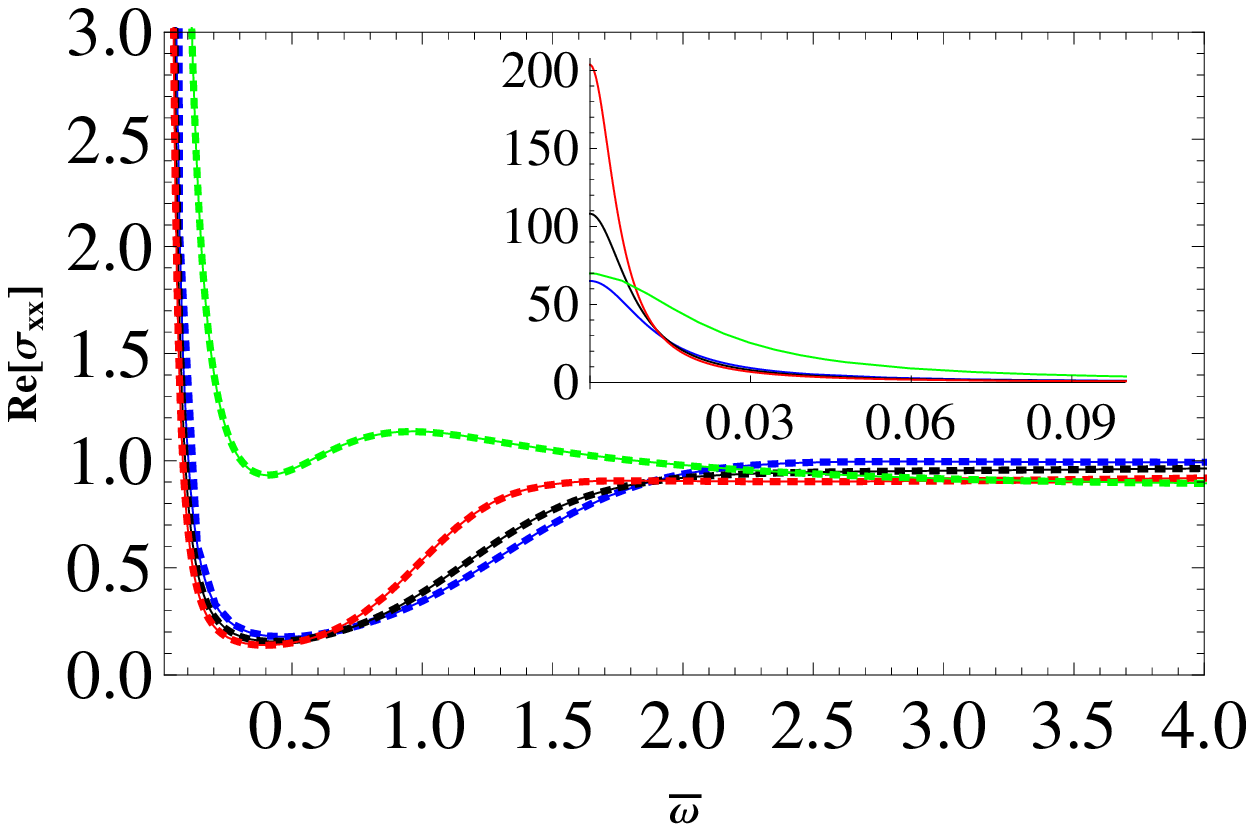}
\hspace{1cm}\includegraphics[width =0.4\textwidth]{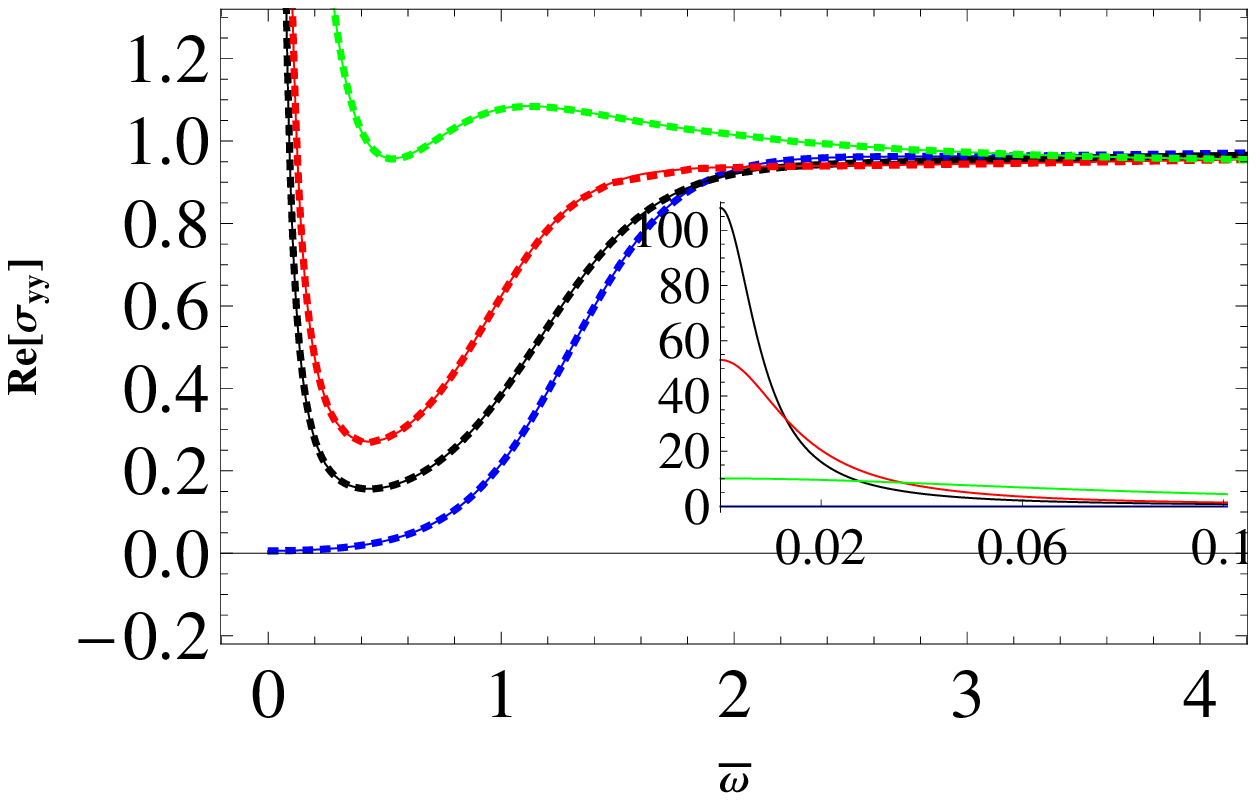}
\includegraphics[width =0.4\textwidth]{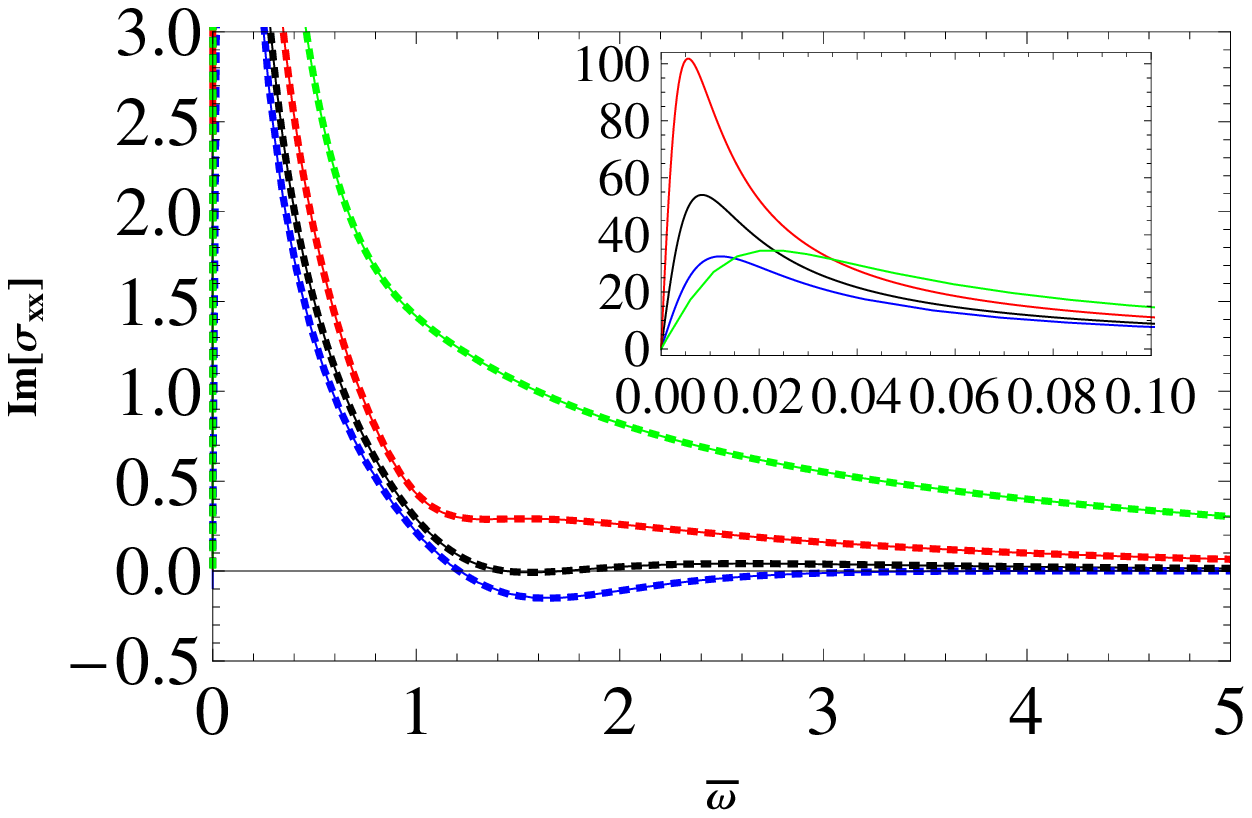}
\hspace{1cm}\includegraphics[width =0.4\textwidth]{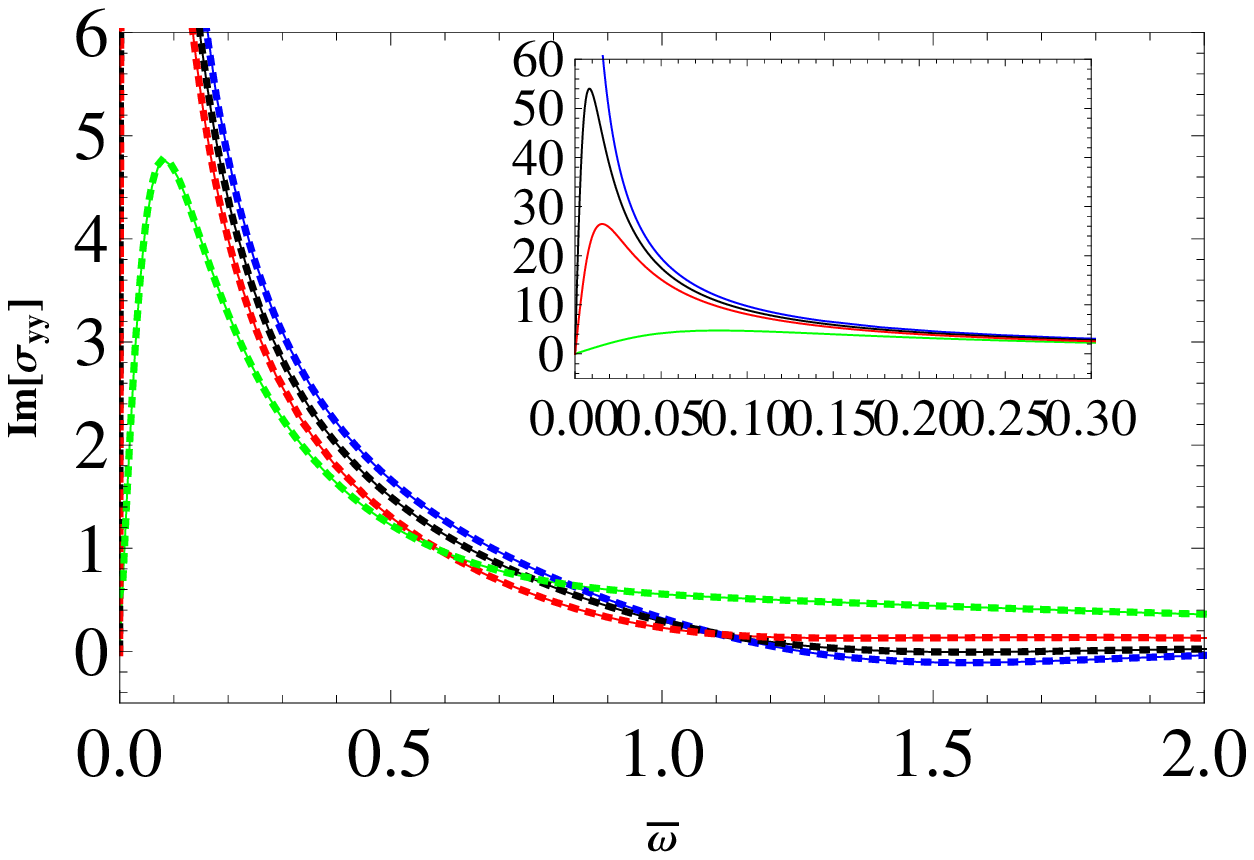}
\vspace{-0.5 cm}
\caption{At a given temperature ($\kappa=1$), electric conductivity with $\alpha_2=0$ (Blue), $2$ (Black),  $4$ (Red) and $6$ (Green). }\label{F:ReImSigma2CasesX}
\end{figure}

\begin{figure}
\vspace{0cm}
\centering
\includegraphics[width =0.4\textwidth]{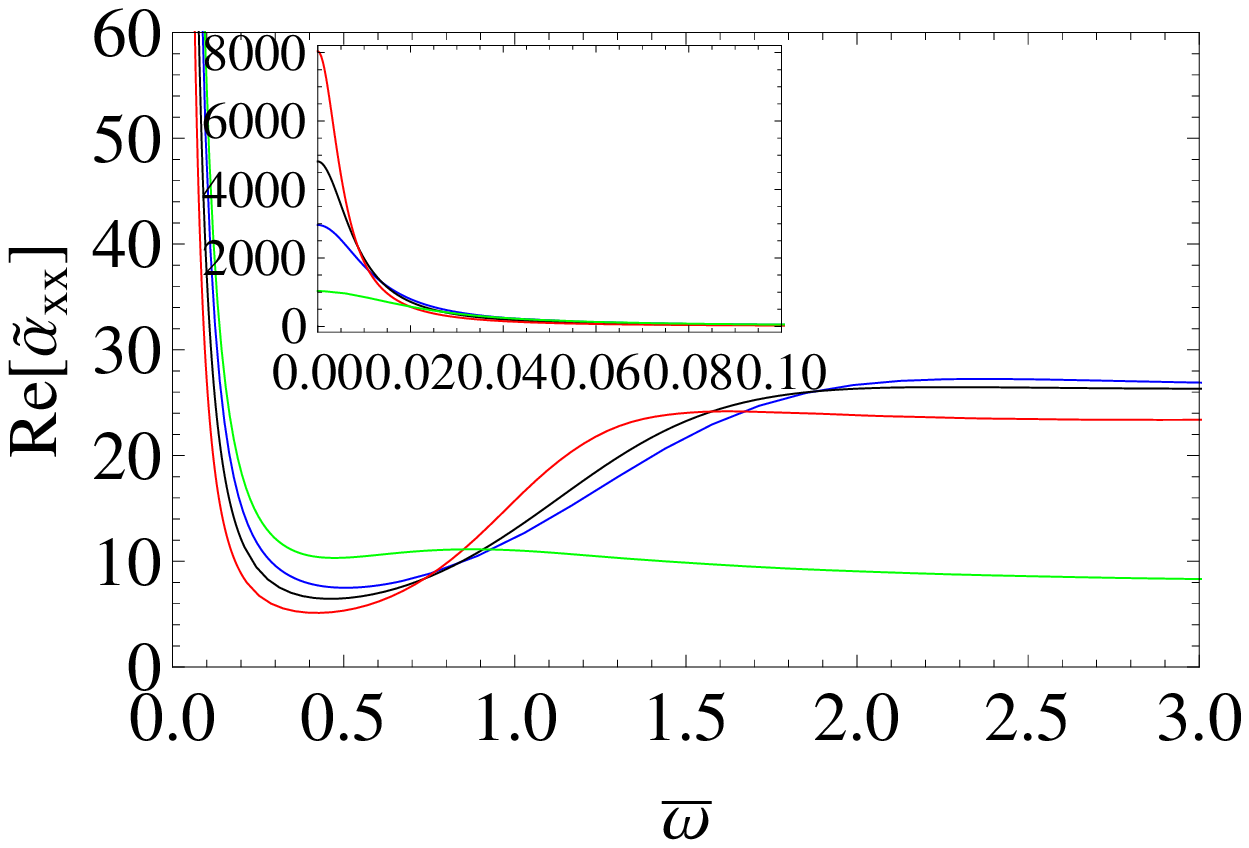}
\hspace{1cm}\includegraphics[width =0.4\textwidth]{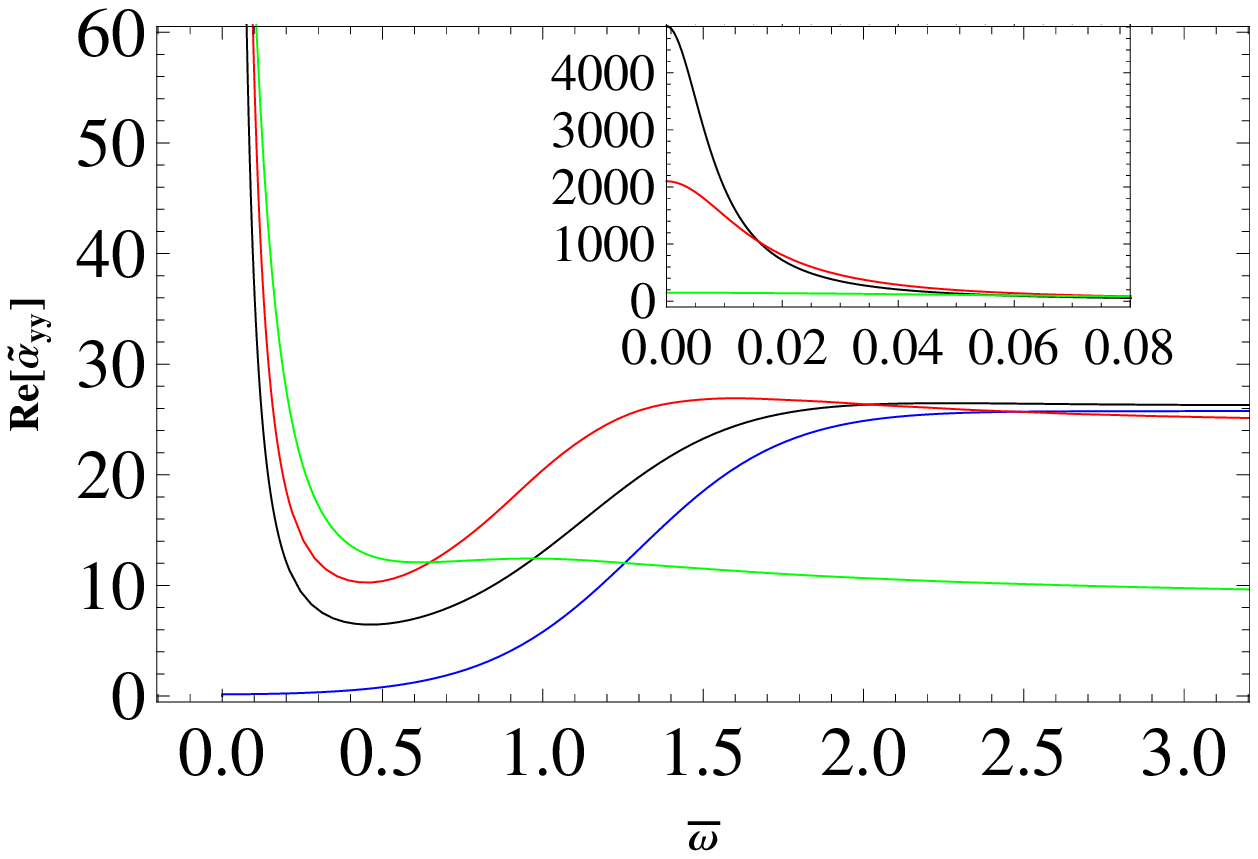}
\includegraphics[width =0.4\textwidth]{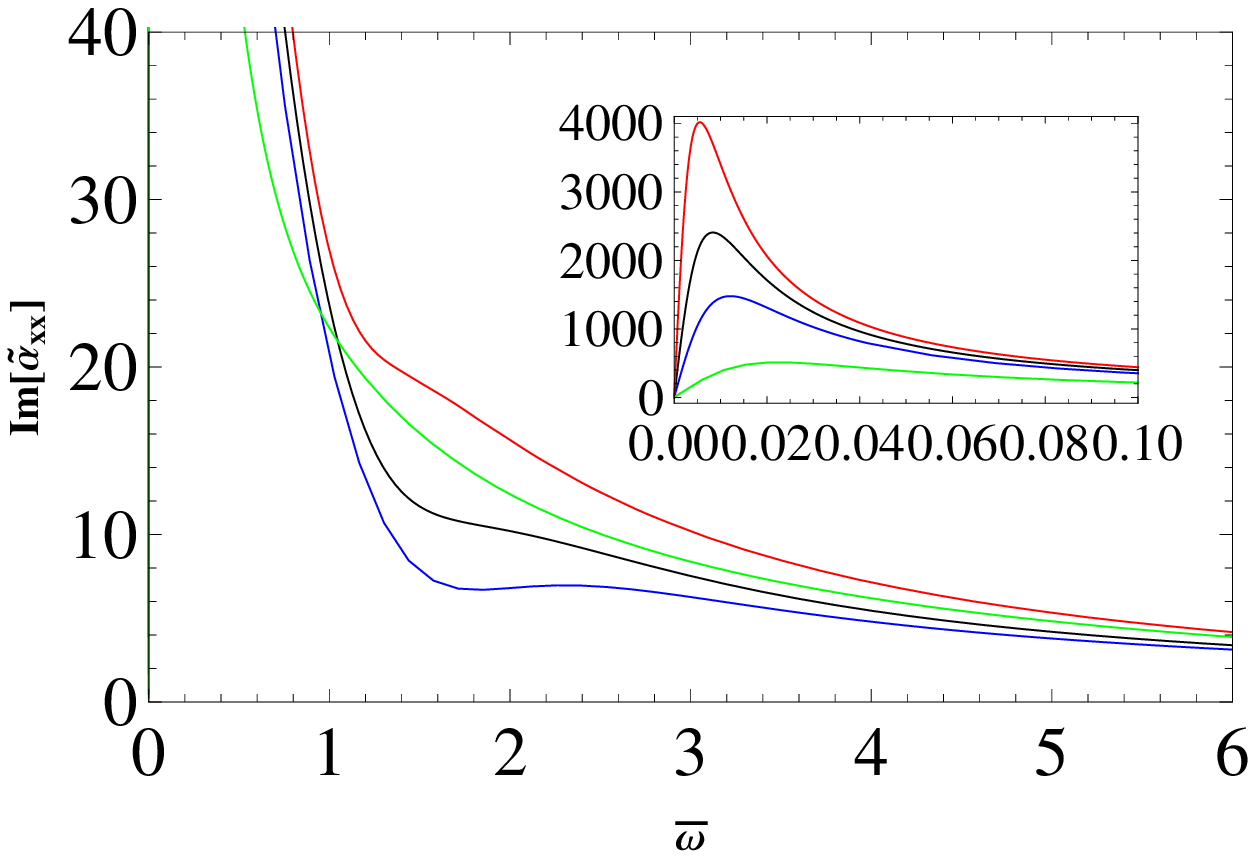}
\hspace{1cm}\includegraphics[width =0.4\textwidth]{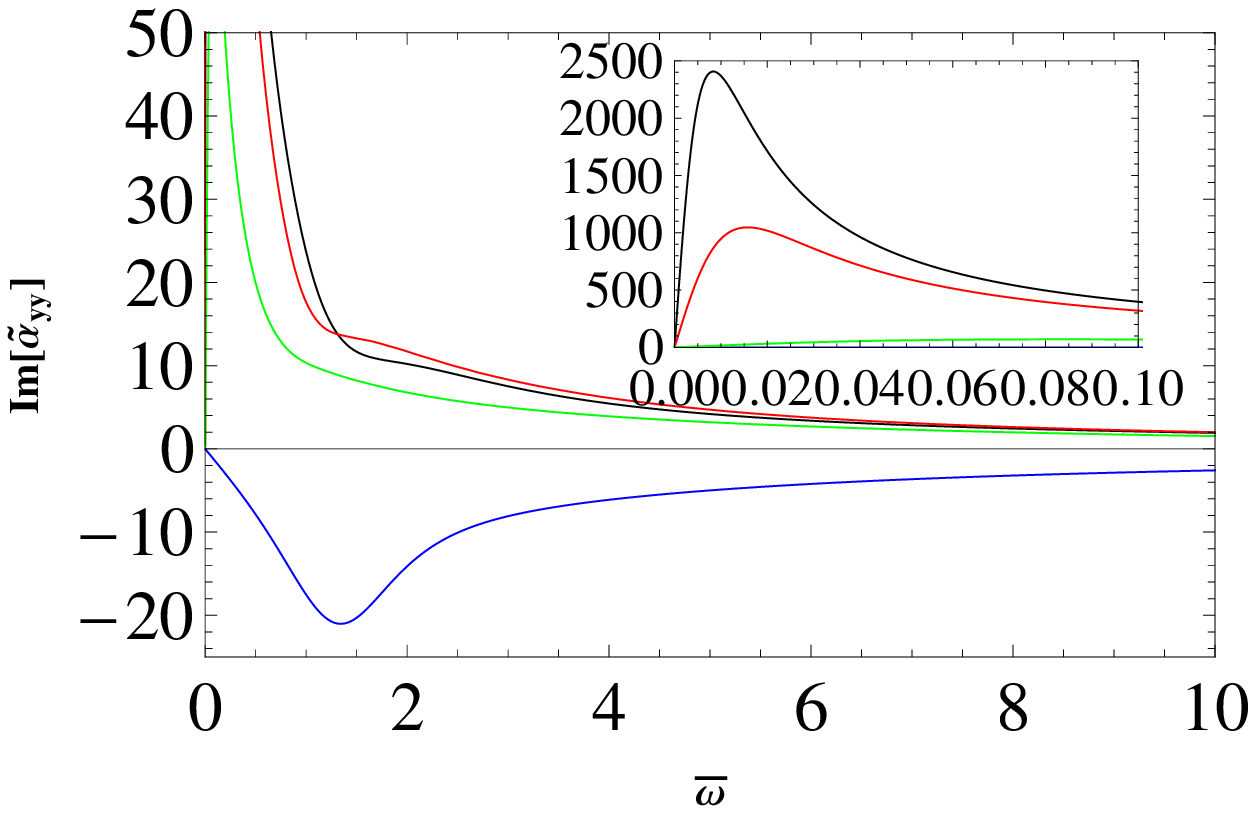}
\vspace{-0.5 cm}
\caption{ Fix $\alpha_1=2$ with $\kappa=1$ and  $\alpha_2=0$ (Blue), $2$ (Black),  $4$ (Red), and $6$ (Green).}\label{F:AlphaKappaXY}
\end{figure}

\begin{figure}
\vspace{-0cm}
\centering
\includegraphics[width =0.4\textwidth]{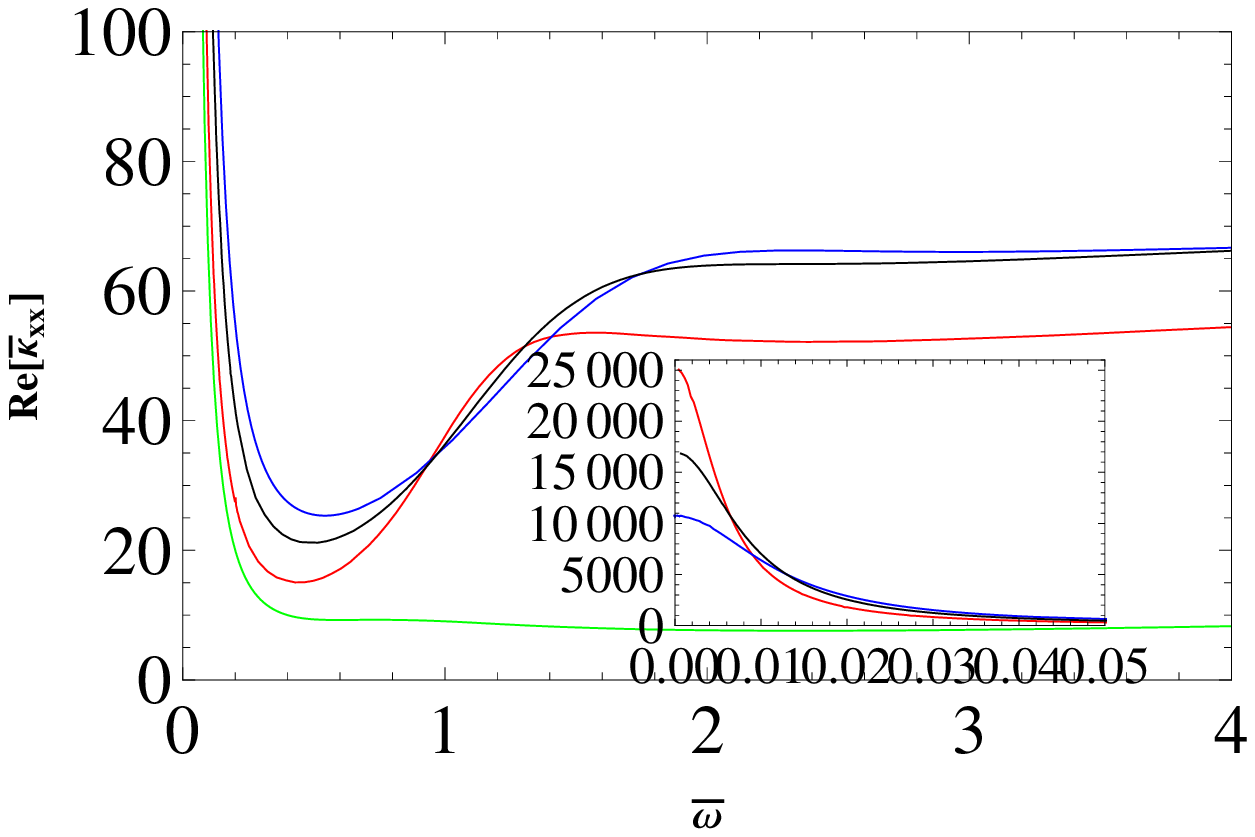}
\hspace{1cm} \includegraphics[width =0.4\textwidth]{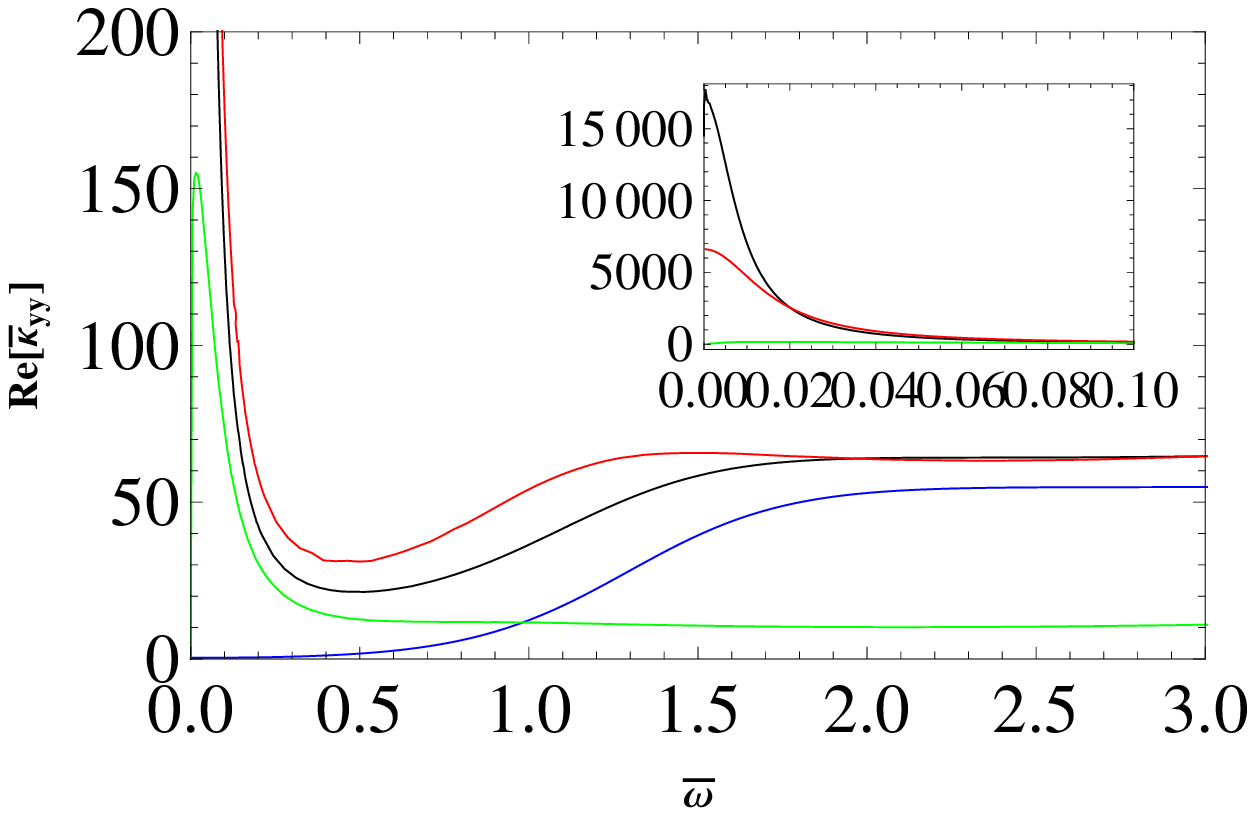}
\includegraphics[width =0.4\textwidth]{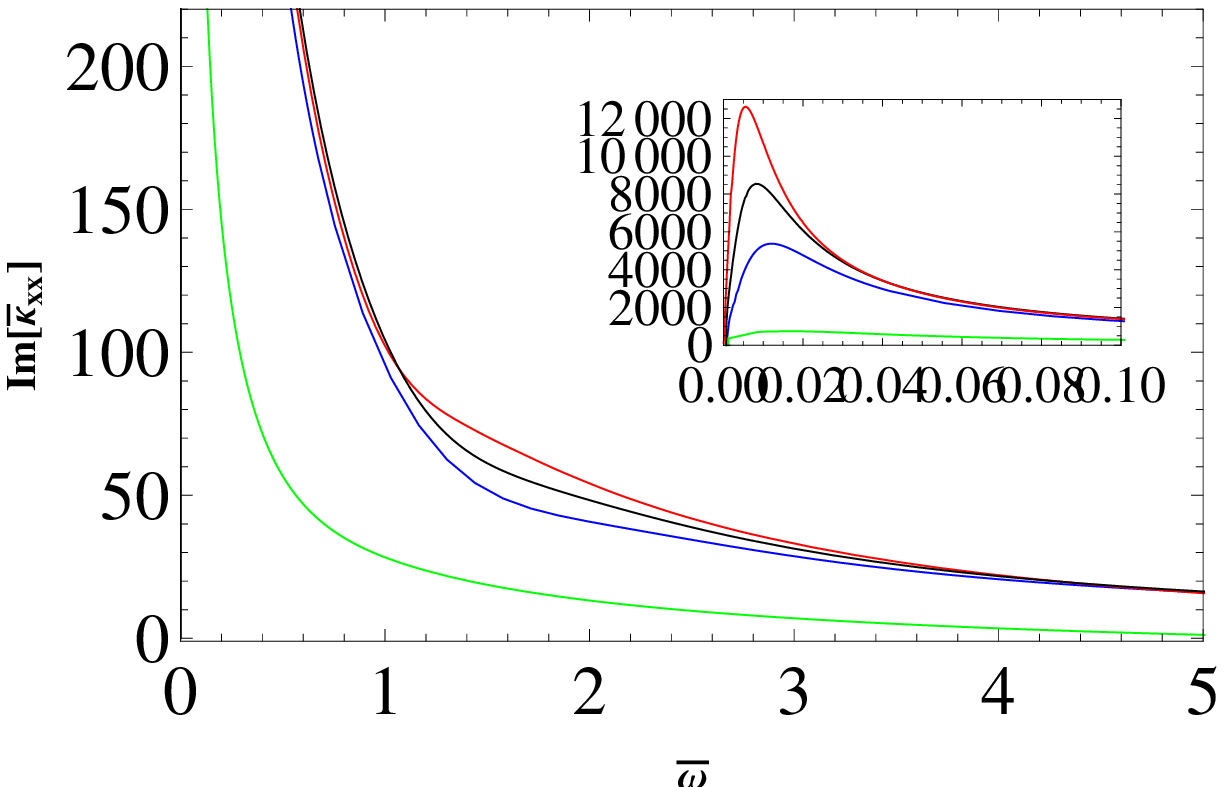}
\hspace{1cm} \includegraphics[width =0.4\textwidth]{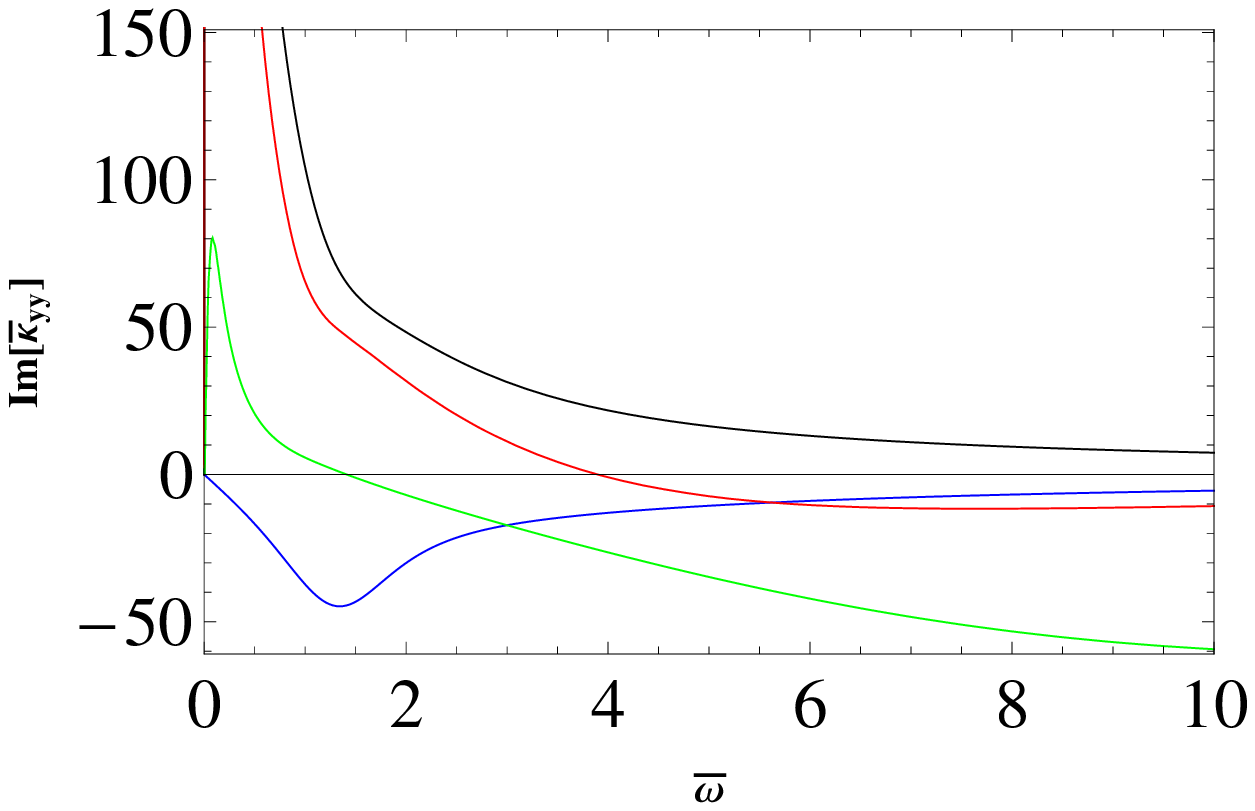}
\vspace{-0.5 cm}
\caption{ Fix $\alpha_1=2$ with $\kappa=1$ and  $\alpha_2=0$ (Blue), $2$ (Black),  $4$ (Red), and $6$ (Green).}\label{F:ReAlphaKappaX}
\end{figure}

For simplicity, let us fix $L=1$ and $\alpha_1=2$ where $\alpha_2$ measures the anisotropy. After the numerical calculation following the previous procedure, we obtain electric, thermoelectric and thermal conductivities depending on the frequency in Fig.  \ref{F:ReImSigma2CasesX},  \ref{F:AlphaKappaXY} and  \ref{F:ReAlphaKappaX} respectively. 
In spite of the fact that there is no explicit coupling between fluctuations in $x$- and $y$-directions. these plots show the nontrivial anisotropy dependence. In other words, even when the $x$-direction momentum relaxation is fixed, the change of the $y$-direction momentum relaxation alters all conductivities in $x$- and $y$-directions. This is because information about momentum relaxations in $x$- and $y$-directions are encoded into the background geometry. Due to this reason, conductivities of this model are sensitive to the anisotropy (see  Fig.  \ref{F:ReImSigma2CasesX},  \ref{F:AlphaKappaXY} and  \ref{F:ReAlphaKappaX}). Intriguingly, the electric conductivity in the high frequency limit seems to converge rapidly into the same value. This implies that the anisotropic effect on the electric conductivity becomes less important in the high frequency regime.

\begin{figure}
\centering
\includegraphics[width =0.45\textwidth]{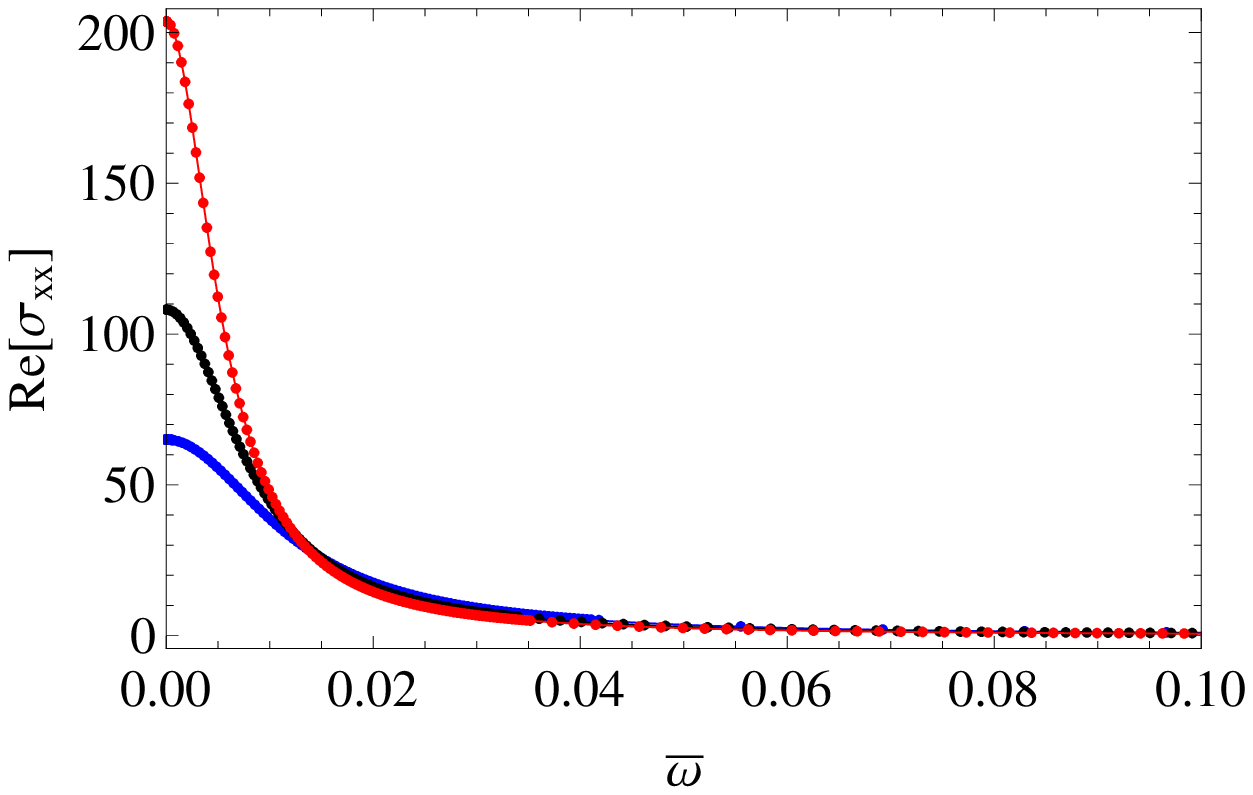}
\hspace{1cm}\includegraphics[width =0.45\textwidth]{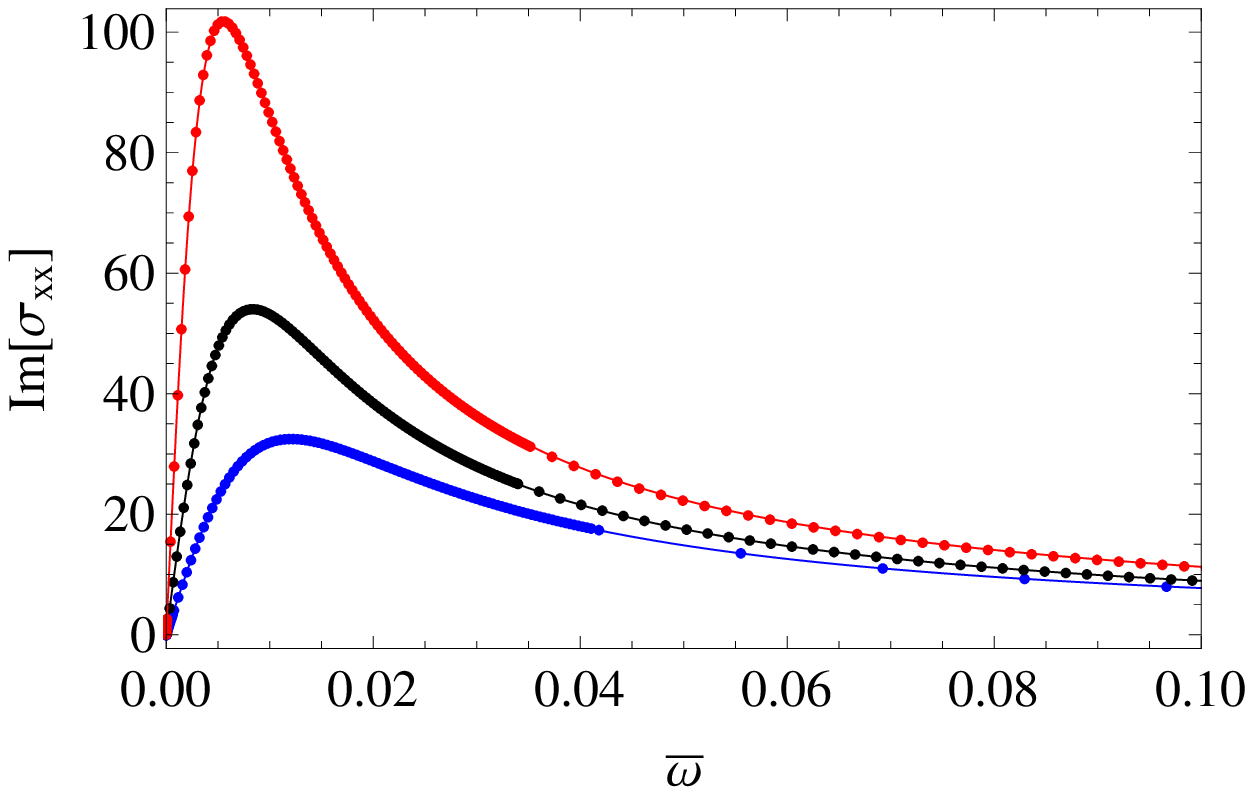}
\caption{Drude peak for $\alpha_1=2$ and $\kappa=1$ with $\alpha_2=0$ (Blue), $2$ (Black) and $4$ (Red). Notice that a solid line indicates the analytic result from the Drude formula, while dots represent the numerical results. }\label{F:SigmaXDrudePowerLaws}
\end{figure}

In the low frequency regime, Fig.  \ref{F:ReImSigma2CasesX},  \ref{F:AlphaKappaXY} and  \ref{F:ReAlphaKappaX} show a Drude-like peak, so that it would be interesting to compare it with the Drude formula 
\begin{align}\label{DrudeForm}
\Gamma= \dfrac{k \tau}{1- i \omega \tau} ,
\end{align}
where $\Gamma$ represents different kinds of conductivity. In general, the relaxation time $\tau$ and coefficient $k$ depend on the momentum relaxation. When $\alpha_1$ is fixed, we can investigate their anisotropy dependence by varying $\alpha_2$. We depict the electric conductivity, $\Gamma=\sigma_{xx}$, together with the result of the Drude formula in Fig. \ref{F:SigmaXDrudePowerLaws}. This result shows that our numerical results are perfectly matched to the Drude formula when we take parameter values in Table 1. When the momentum relaxation in $y$-direction increases, the relaxation time and coefficient $k$ in $x$-direction also increase. In addition, the Drude peak becomes narrow 
as $\alpha_2$ increases within the range, $0 \le\alpha_2  \le 4$, similar to \cite{Koga:2014hwa}.

\vspace{0.5cm}
\begin{center}
\begin{tabular}{ | l | l | l | l |  l |  l  |}
\hline
$\sigma_{xx}$             & $\gamma$ & $b$ & $c$ & $k$ & $ \tau$ \\
\hline
$\alpha_2=0$            &2/3 & 2.88 & -5.65 & 0.785 & 82.7 \\     \hline
$\alpha_2=2$            & 0.89 & 1.3 & -0.8 & 0.9 & 120  \\      \hline
$\alpha_2=4$            & 0.626 & 6.87 & -24.5 & 1.13 & 180    \\     \hline  
\end{tabular} \\
\end{center}
Table 1.  Parameters of the Drude formula fitting the electric conductivity well. 
\vspace{0.5cm}

\begin{figure}
\centering
\includegraphics[width =0.45\textwidth]{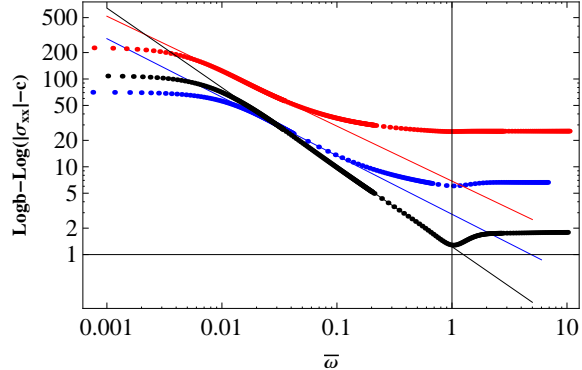}
\caption{ The magnitude of the electric conductivity for $\alpha_1=2$ and $\kappa=1$ with $\alpha_2=0$ (Blue), $2$ (Black) and $4$ (Red). The slope of the straight line denotes the power law of the electric conductivity.}\label{F:SigmaXDrudePowerLaws1}
\end{figure}

Now, let us investigate the magnitude of the electric conductivity in an intermediate frequency regime \cite{Ling:2013nxa}. Using the previous numerical result, it is plotted in Fig. \ref{F:SigmaXDrudePowerLaws1}, which shows a specific scaling behavior in an intermediate frequency regime. To clarify the scaling behavior, we consider the following power law behavior
\begin{align}\label{PowerLaws}
|\sigma| = \dfrac{b}{\omega^\gamma}+c
\end{align}
where the exponent $\gamma$ determines the scaling behavior. In order to fit the numerical data, we found best parameter values in Table 1. In the isotropic case, the scaling exponent is given by $\gamma= 0.89$. When the anisotropy becomes large, our result shows that the scaling exponent decreases.

\begin{figure}
\centering
\includegraphics[width =0.4\textwidth]{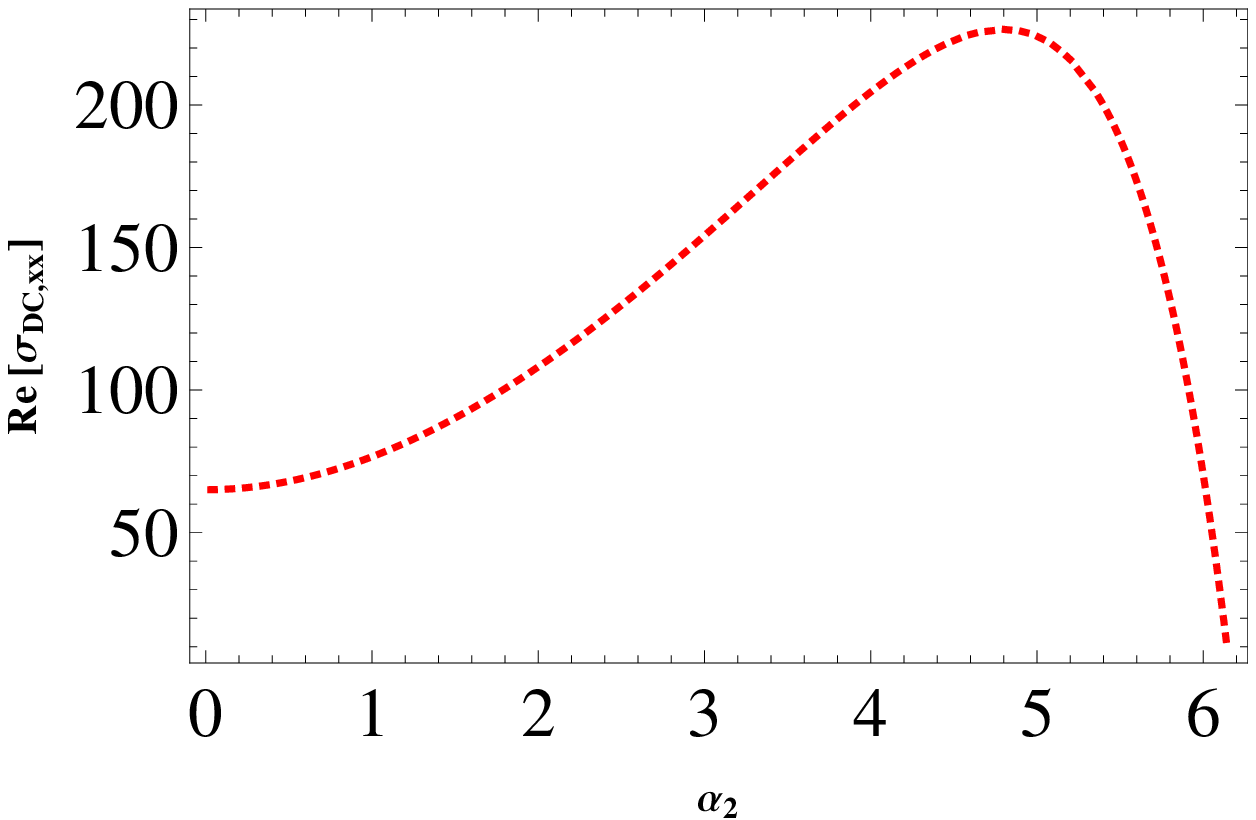}
\hspace{1cm} \includegraphics[width =0.44\textwidth]{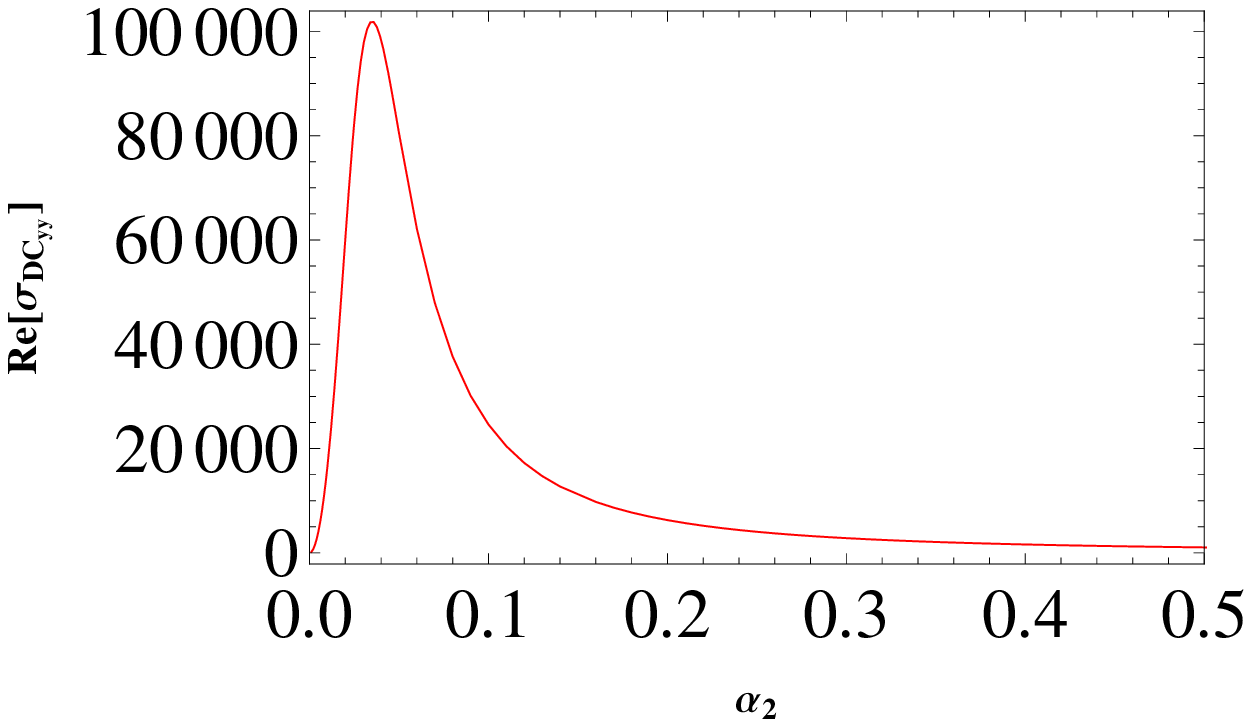}
\includegraphics[width =0.4\textwidth]{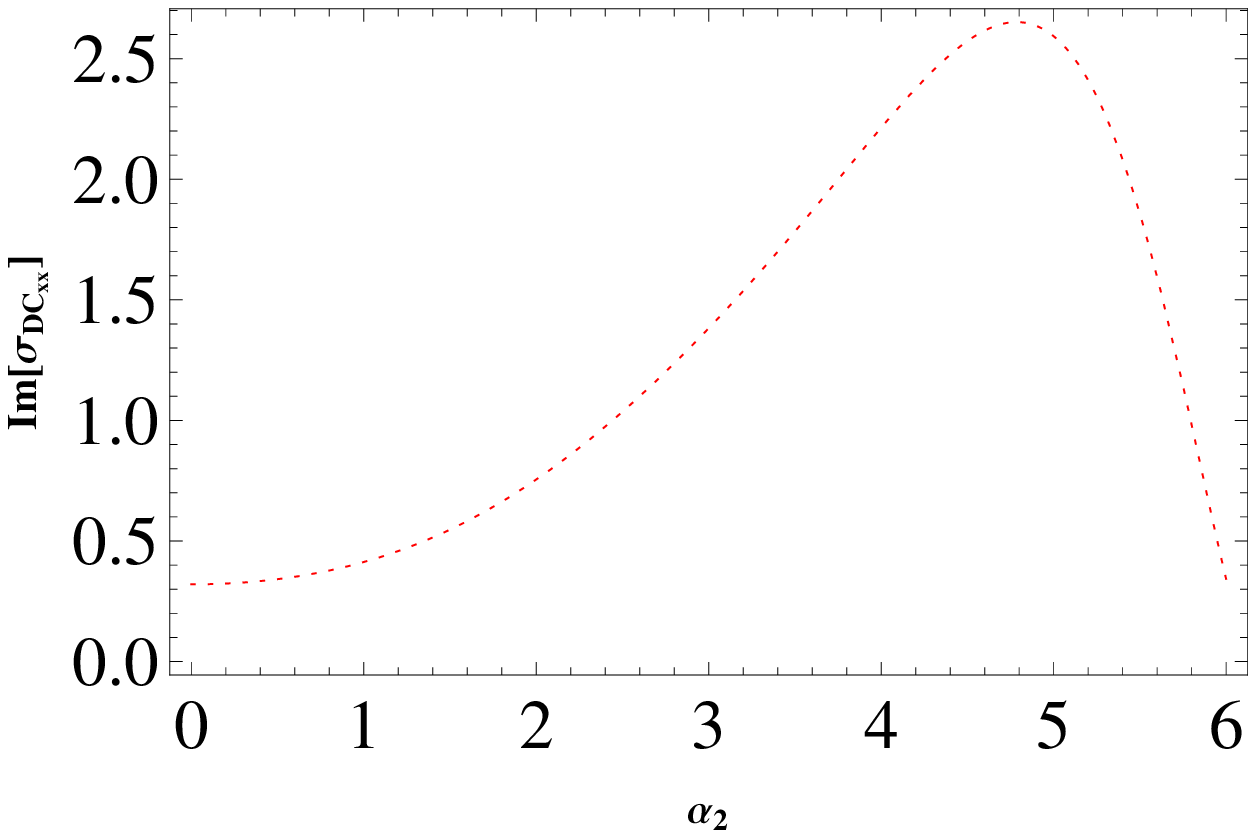}
\hspace{1cm} \includegraphics[width =0.44\textwidth]{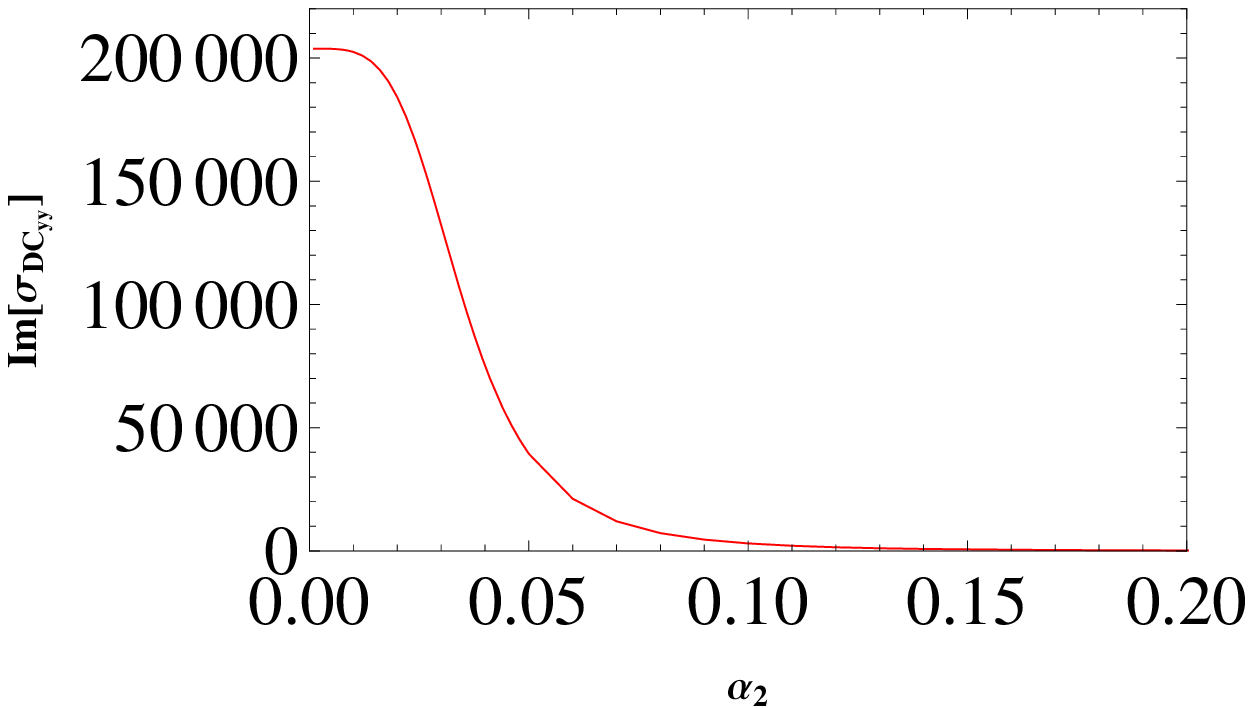} \\
\vspace{-0.0cm}
\caption{DC conductivities with $\alpha_1=2$ and $\kappa=1$. The range of $\alpha_2$ is restricted to $0<\alpha_2<6.1393$. }\label{F:ReSigmaDC}
\end{figure}

\begin{figure}
\centering
\includegraphics[width =0.5\textwidth]{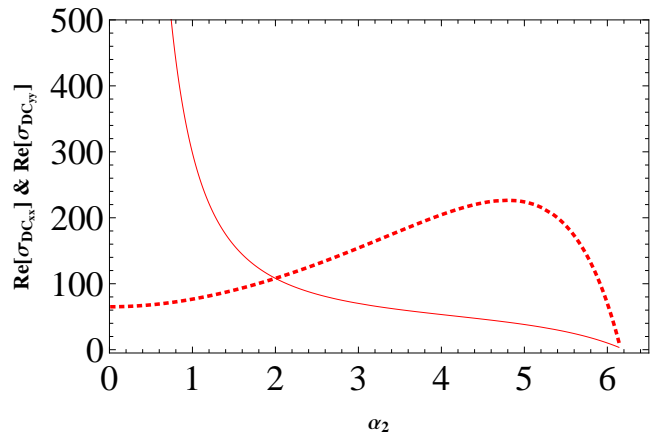}
\caption{ $\alpha_2$ dependence of DC conductivities, $\sigma_{xx}$ (dotted) and $\sigma_{yy}$ (solid), with $\alpha_1=2$ and $\kappa=1$.}\label{F:ImSigmaDC}
\end{figure}

Finally, let us study how the anisotropy affects the DC conductivity. As shown in Fig. \ref{F:ReImSigma2CasesX}, the low frequency behavior of the electric conductivity is sensitive to the anisotropy. We plot the DC conductivity depending on the anisotropy in Fig. \ref{F:ReSigmaDC} and its real part in Fig. \ref{F:ImSigmaDC}. 
In Fig. \ref{F:ReSigmaDC}, one can see that there exist maximum values for $x$- and $y$-direction DC conductivities at certain critical frequencies. Below these critical frequencies $x$- and $y$-direction DC conductivities increase, whereas they decrease above the critical frequencies. In Fig. \ref{F:ImSigmaDC}, one can see that the $x$-direction DC conductivity is the same as that of $y$-direction at $\alpha_2=2$. This is the consistent result because the isotropy is restored at this point. 
Before closing this section, it should be noticed that there seems to be an upper bound for $\alpha_2$. When we solved equations for the background geometry with $\alpha_1=2$, we failed to find a numerical solution above $\alpha_2 \approx 6.1393$. If there is no such an upper bound, the extrapolation in Fig. \ref{F:ImSigmaDC} says that the conductivity can change its sign at a certain value of $\alpha_2$. However, the upper bound we found does not allow the regime of a negative conductivity.

\section{Conclusion \label{conclusion} }

In condensed matter experiments, anisotropy is one of important ingredients of material. However, it is not generally easy to understand the anisotropic effect in the strongly interacting system. In this work, we tried to figure out qualitative features of the anisotropy by using the AdS/CFT correspondence. In order to mimic such an anisotropy, we considered an Einstein-Maxwell-dilaton-axion Model as a dual gravity theory. In this model, the gauge and dilaton fields describe matter and a nontrivial coupling of the dual field theory. On the other hand, the momentum relaxations denoted by $\alpha_1$ and $\alpha_2$ were introduced to represent breaking of the translational symmetry in $x$- and $y$-directions. Taking different values for $\alpha_1$ and $\alpha_2$ further breaks the rotational symmetry which is the origin of the anisotropy. By solving Einstein equations, we constructed an anisotropic charged black hole solution numerically. Furthermore, we took into account dynamics of vector fluctuations on this charged black hole which allows us to investigate the effect of the anisotropy on the transport coefficients like electric, thermoelectric and thermal conductivities.  
 There are several remarkable points for the linear responses in an anisotropic medium. 
\begin{itemize}
\item  When a momentum relaxation is turned on, we numerically showed that conductivities in $x$- as well as $y$-directions become finite as expected. 

\item Although equations for vector fluctuations in $x$- and $y$-directions are not coupled, we found that the $y$-direction momentum relaxation can affect both $x$- and $y$-direction linear responses. On the gravity side, it is because the background geometry involves information about $x$- and $y$-direction momentum relaxations.

\item There exists a critical momentum relaxation at which the DC conductivity has a maximum value. 

\item There seems to be an upper bound of the anisotropy above which the dual geometry does not exist. This upper bound does not allow the sign change of the DC conductivity.

\item In the low frequency regime, the electric conductivity shows a Drude peak. When the $x$-direction momentum relaxation is fixed to be $\alpha_1=2$, the Drude peak becomes broader as the $y$-direction momentum relaxation increases.

\item In the intermediate frequency regime, the magnitude of the electric conductivity shows a specific scaling behavior. Comparing it with the power law behavior, our results show that the critical exponent becomes smaller as the anisotropy increases.

\end{itemize}

\vspace{1cm}

{\bf Acknowledgement} \\

S. Khimphun acknowledges the Korea Ministry of Science, ICT and Future Planning for the support of the Visitors Program at the Asia Pacific Center for Theoretical Physics (APCTP). 
We were supported by the Korea Ministry of Education, Science and Technology, Gyeongsangbuk-Do and Pohang City.
B.-H. Lee was supported by the National Research Foundation of Korea (NRF) grant funded by the Korea government (MSIP) (2014R1A2A1A01002306). C. Park was supported by Basic Science Research Program through the National Research Foundation of Korea (NRF) funded by the Ministry of Education (NRF-2013R1A1A2A10057490).

\end{document}